\documentclass[conference ]{IEEEtran}
\def\BibTeX{{\rm B\kern-.05em{\sc i\kern-.025em b}\kern-.08em
    T\kern-.1667em\lower.7ex\hbox{E}\kern-.125emX}}

\usepackage[normalem]{ulem}
\usepackage{upgreek}
\usepackage{subfig}
\usepackage{color,xcolor}
\usepackage{xspace}
\usepackage{multirow}
\usepackage{booktabs}
\usepackage{lipsum}
\usepackage{soulutf8}
\usepackage{diagbox}
\usepackage{physics}

\usepackage{graphicx}
\usepackage{colortbl}

\usepackage{enumitem}
\usepackage{braket}
\usepackage{calc}

\usepackage{amsmath}
\usepackage{amsfonts}
\usepackage{url}

\usepackage{afterpage}
\usepackage{cancel}

\usepackage[ruled,vlined]{algorithm2e}
\usepackage{setspace}

\usepackage{pifont}

\usepackage{titlesec}

\usepackage{hyperref}
\hypersetup{
    colorlinks=true,
    linkcolor=magenta,
    filecolor=magenta,      
    urlcolor=blue,
}

\usepackage{xcolor}
\usepackage{soul}

\definecolor{citecolor}{RGB}{34,139,34}
\definecolor{mydarkblue}{rgb}{0,0.08,1}
\definecolor{mydarkgreen}{rgb}{0.02,0.6,0.02}
\definecolor{mydarkred}{rgb}{0.8,0.02,0.02}
\definecolor{mydarkorange}{rgb}{0.40,0.2,0.02}
\definecolor{mypurple}{RGB}{111,0,255}
\definecolor{myred}{rgb}{1.0,0.0,0.0}
\definecolor{mygold}{rgb}{0.75,0.6,0.12}
\definecolor{myblue}{rgb}{0,0.2,0.8}
\definecolor{mydarkgray}{rgb}{0.,0.2,0.2}

\definecolor{lightred}{RGB}{255,235,235}
\definecolor{lightgreen}{RGB}{235,255,235}
\definecolor{lightblue}{RGB}{235,235,255}
\definecolor{lightcyan}{RGB}{235,255,255}
\definecolor{lightmagenta}{RGB}{255,235,255}
\definecolor{lightyellow}{RGB}{255,255,235}

\definecolor{qxkcolor}{RGB}{215,235,255}
\definecolor{softmaxcolor}{RGB}{230,235,255}
\definecolor{probxvcolor}{RGB}{255,255,235}

\definecolor{topkcolor}{RGB}{255,235,235}
\definecolor{zecolor}{RGB}{255,255,235}
\definecolor{dynacolor}{RGB}{235,255,255}

\definecolor{reviewcolor}{RGB}{0,0,200}

\renewcommand\footnotemark{}

\newcommand{\name}{Atomique\xspace}

\newcommand{\fpqas}{FPQAs\xspace}
\newcommand{\raa}{RAA\xspace}
\newcommand{\raas}{RAAs\xspace}

\newcommand{\x}{$\times$\xspace}

\newcommand{\etal}{\emph{et al.}\xspace}

\newcommand{\cnot}{\texttt{CNOT}}

\newcommand{\faas}{FAAs\xspace}

\newcommand{\zz}{\texttt{ZZ}}
\newcommand{\cz}{\texttt{CZ}}

\newcommand{\graphname}{complete multipartite coupling graph}

\newcommand{\fpqafull}{\textit{field programmable qubit arrays}\xspace}
\newcommand{\raafull}{\textit{reconfigurable atom arrays}\xspace}

\newcounter{rlabelno}

\definecolor{color1}{rgb}{1, 1, 0.9}
\definecolor{color2}{rgb}{1, 0.9, 1}
\definecolor{color3}{rgb}{0.9, 1, 1}

\definecolor{color4}{rgb}{1, 0.9, 0.9}
\definecolor{color5}{rgb}{0.9, 0.9, 1}
\definecolor{color6}{rgb}{0.9, 1, 0.9}

\definecolor{color7}{rgb}{0.8, 0.9, 1}
\definecolor{color8}{rgb}{0.9, 1, 0.8}
\definecolor{color9}{rgb}{1, 0.8, 0.9}

\newcolumntype{a}{>{\columncolor{color1}}c}
\newcolumntype{b}{>{\columncolor{color2}}c}
\newcolumntype{d}{>{\columncolor{color3}}c}
\newcolumntype{e}{>{\columncolor{color4}}c}
\newcolumntype{f}{>{\columncolor{color5}}c}
\newcolumntype{g}{>{\columncolor{color6}}c}
\newcolumntype{h}{>{\columncolor{color7}}c}
\newcolumntype{i}{>{\columncolor{color8}}c}
\newcolumntype{j}{>{\columncolor{color9}}c}

\usepackage{listings} 
\usepackage{xcolor} 

\definecolor{codegray}{rgb}{0.5,0.5,0.5}
\definecolor{codegreen}{rgb}{0,0.6,0}
\definecolor{codepurple}{rgb}{0.58,0,0.82}
\definecolor{backcolour}{rgb}{0.95,0.95,0.92}

\usepackage{hyperref}

\author{Hanrui Wang$^{1,2}$, Pengyu Liu$^3$, Daniel Bochen Tan$^2$, Yilian Liu$^4$, Jiaqi Gu$^5$\\David Z. Pan$^6$, Jason Cong$^2$, Umut A. Acar$^3$, Song Han$^1$\\ \footnotesize $^1$Massachusetts Institute of Technology, 
$^2$University of California, Los Angeles, 
$^3$Carnegie Mellon University, $^4$Cornell University\\$^5$Arizona State University, $^6$University of Texas at Austin}

\usepackage{titlesec}
\usepackage{fancyhdr}

\begin{document}

\title{Atomique: A Quantum Compiler for \\Reconfigurable Neutral Atom Arrays}

\date{}
\maketitle
\pagestyle{plain}
\thispagestyle{fancy}
\renewcommand\sectionmark[1]{} 

\begin{abstract}

The neutral atom array has gained prominence in quantum computing for its scalability and operation fidelity.
Previous works focus on \textit{fixed atom arrays (\faas)} that require extensive SWAP operations for long-range interactions. This work explores a novel architecture \raafull (\raas), also known as \fpqafull (\fpqas), which allows for coherent atom movements during circuit execution under some constraints.  
Such atom movements, which are unique to this architecture, could reduce the cost of long-range interactions significantly if the atom movements could be scheduled strategically.

In this work, we introduce \name, a compilation framework designed for qubit mapping, atom movement, and gate scheduling for \raa. \name contains a \textit{qubit-array mapper} to decide the coarse-grained mapping of the qubits to arrays, 
leveraging MAX k-Cut on a constructed gate frequency graph to minimize SWAP overhead. 
Subsequently, a \textit{qubit-atom mapper} determines the fine-grained mapping of qubits to specific atoms in the array and considers load balance to prevent hardware constraint violations. 
We further propose a router that identifies parallel gates, schedules them simultaneously, and reduces depth.

We evaluate \name across 20+ diverse benchmarks, including generic circuits (arbitrary, QASMBench, SupermarQ), quantum simulation, and QAOA circuits. 
\name consistently outperforms IBM Superconducting, FAA with long-range gates, and FAA with rectangular and triangular topologies, achieving significant reductions in depth and the number of two-qubit gates.

\end{abstract}


\section{Introduction}

The landscape in the field of quantum computing is rapidly evolving. Alongside the development of superconducting systems featuring up to 433 qubits~\cite{ibm433, ibm127, rigetti, google72, intel49, krantz2019quantum}, 
the field of neutral atom arrays is also making remarkable strides, in which the architecture of each atom can be used as a qubit. Infleqtion recently announced a groundbreaking 1600 qubit system~\cite{Infleqtion_2024}; 
Ebadi et al.\mbox{~\cite{ebadi_na}} showcased a system with up to 289 qubits; QuEra introduced a 256-qubit system available on AWS\mbox{~\cite{wurtz2023aquila, evered2023highfidelity,quera256}};
Pasqal has unveiled a device with 324 qubits\mbox{~\cite{Schymik2022pasqal}}; Atom Computing has developed technology supporting over 1000 qubit traps\mbox{~\cite{Barnes2022atom, norcia2024iterative}}; and Bluvstein et al.\mbox{~\cite{bluvstein2023logical}} also demonstrate implementation of various error correction codes with neutral atom arrays.

\begin{figure}[t]
    \centering
    \includegraphics[width=\columnwidth]{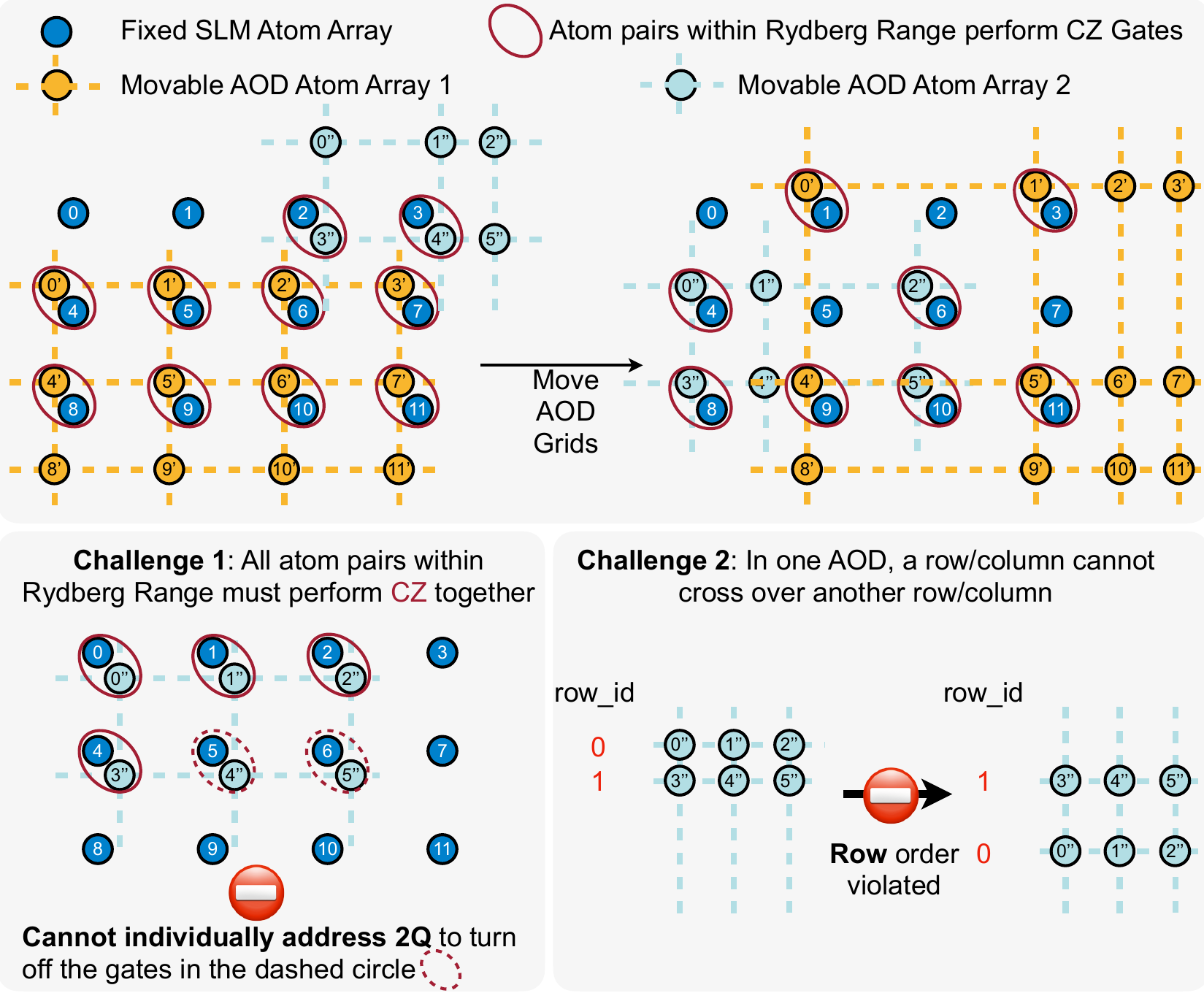}
    \vspace{-10pt}
    \caption{Reconfigurable neutral atom array with fixed and movable atoms. 
    }
    \label{fig:teaser}
    \vspace{-15pt}
\end{figure}

 This momentum has catalyzed the architecture community to focus on compiler development for these platforms that map and route qubits~\cite{li2019tackling, lao20222qan,isca19-murali-linke-martonisi-abhari-nguyen-alderete-triq-architecture-studies, DBLP:conf/isca/SmithT19, DBLP:conf/asplos/DuckeringBLC21, DBLP:conf/micro/AlamAG20, gokhale2020optimized, DBLP:conf/sc/HuaWLPLZSDZH023}.
Baker et al.~\cite{baker2021exploiting} pioneered the first compiler framework tailored for neutral atom arrays, adapting qubit mapping strategies from superconducting systems and accommodating the unique advantages and constraints of neutral atoms, such as long-range interaction zones and sporadic atom loss. Building on this, Patel et al. (Geyser)~\cite{geyser} proposed enhancing the system's efficiency by resynthesizing circuits into three-qubit blocks to utilize native multiqubit operations.

Although much of the existing research has focused on \textit{ fixed atom arrays (FAA)}, a more versatile architecture has surfaced: the \raafull (\raas), which is also known as \fpqafull (\fpqas).
This architecture allows two atoms (qubits) to communicate if they are within the Rydberg radius.
Furthermore, the architecture offers fixed atoms, realized using spatial light modulators (SLM), and movable atoms, realized by acousto-optic deflectors (AOD). The AOD atoms can be moved (programmatically) to communicate with other atoms~\cite{bluvstein2022quantum, tan2022},
introducing a dynamic coupling map between qubits determined by atom locations.
This atom mobility makes it possible for otherwise distant atoms to communicate instead of performing extensive SWAP gates as in superconducting architectures.  
\raa also allows us to configure the number of qubits and their locations before execution and move atoms during execution, which is the reason why this architecture is given the name: field programmable qubit arrays.

Fig.~\ref{fig:teaser} illustrates an example with two AOD arrays (represented by yellow and cyan dots) and one SLM array (represented by blue dots). The top subfigure shows a moving step during execution, in which we move the qubits so that the pairs within the Rydberg radius change allowing us to perform different two-qubit gates. Meanwhile, the movement must respect several constraints that the hardware imposes, as depicted in the lower sub-figures. Sec.~\ref{sec:background} discusses more details about atom movement.

\begin{figure}[t]
    \centering
    \includegraphics[width=\columnwidth]{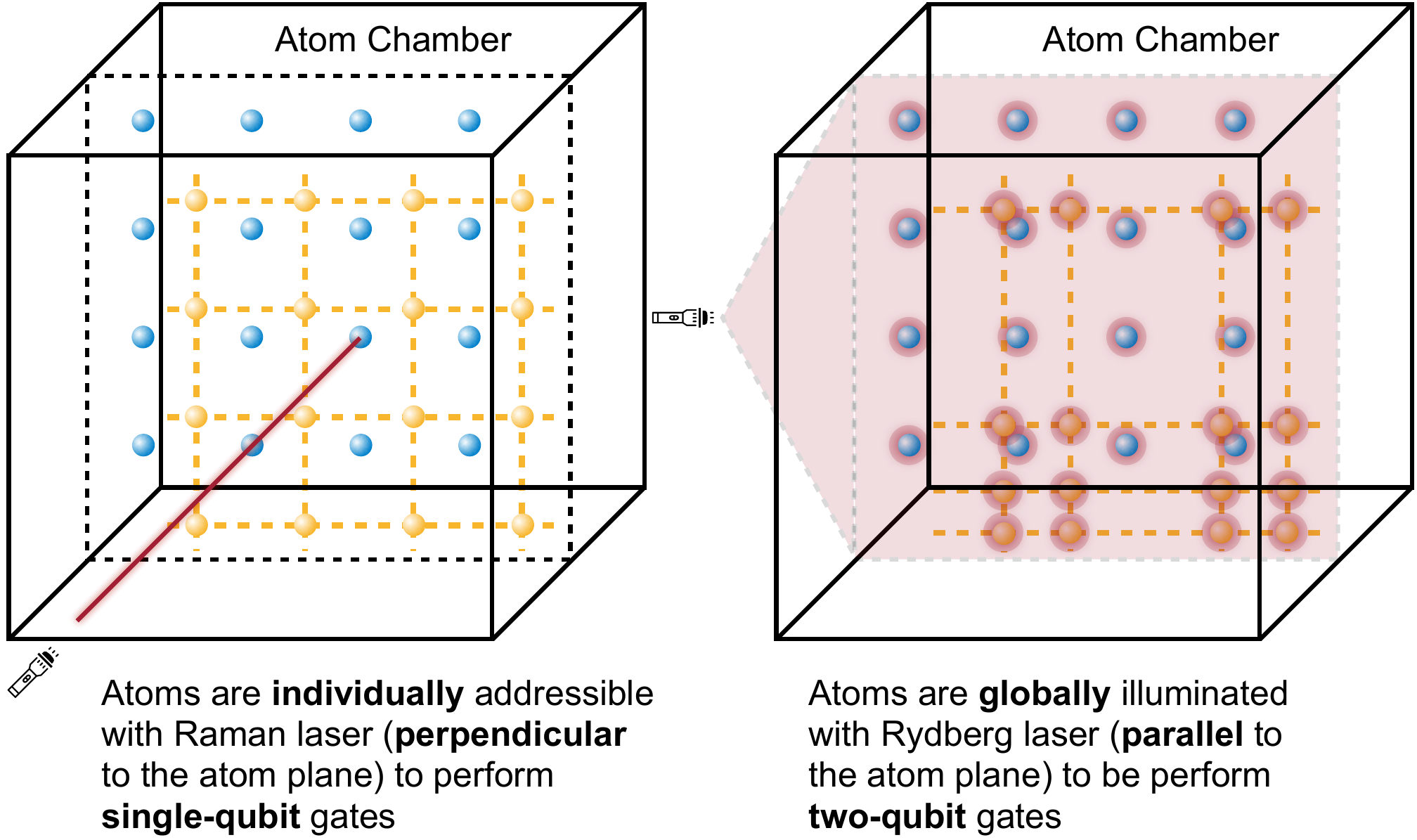}
    \caption{RAA single-qubit gates and two-qubit gates.}    \label{fig:device_illustration}
    \vspace{-20pt}
\end{figure}

As shown in Fig.\mbox{~\ref{fig:device_illustration}}, one-qubit gates can be performed directly using lasers that target atoms from the front, while for two-qubit gates, atoms are moved next to each other and then activated using a laser light from the side, known as a \textit{Rydberg laser}. Importantly, the Rydberg laser affects all atom pairs that are close together, automatically causing them to interact with each other through two-qubit gates. The main takeaway is that while this leads to high-parallel execution, it also means that we have to be very strategic about moving the atoms. We need to ensure that only the desired pairs of atoms are brought close for interaction, to prevent accidental operations between the wrong pairs.
Tan et al.~\cite{tan2022} presented the first \raa compiler, leveraging a satisfiability modulo theories (SMT) solver for qubit mapping and routing. 
Despite its ingenuity, the method faces scalability challenges due to the exponential time complexity of SMT solvers. 
Although it was enhanced with an iterative peeling heuristic~\cite{tan2023compiling}, it still has difficulty to scale beyond 100 qubits. 
Furthermore, it neglects the detrimental impact of atom transfers between the SLM and AOD arrays; such atom transfers can lead to atom loss~\cite{ebadi_na}. 
The consideration of possible atom losses can become significant in iterative algorithms like QAOA or trotterized quantum simulations, leading to potential circuit execution failures.

In this paper, we introduce a scalable compilation framework for \raa, accounting for errors due to atom movements and transfers with additional support for multiple AOD arrays and varying array sizes.
Our approach conceptualizes the hardware as a \textit{\graphname \xspace}, which represents all potential qubit couplings between movable array positions. We partition the execution of circuits into stages. In each stage, we select a subset of edges according to the target circuit to activate the two-qubit gates. Using this representation, we are able to reason about the best schedule for atom movements and laser activations.

This framework addresses two key challenges: 
\begin{itemize}
    \item The absence of individual addressability for two-qubit gates, requiring that the selected edge set must only contain atom pairs that need a two-qubit gate.
    \item The AOD arrays' row/column orders must be preserved, which may forbid two edges being in the same set.
\end{itemize}

To efficiently address these compilation challenges, we employ a two-tier hierarchical mapping strategy. The strategy divides the mapping problem into two separate problems: (1) mapping groups of qubits to arrays and (2) mapping qubits within an array to atoms in order to maximize the parallel execution of gates without violating the hardware constraints.

Firstly, the coarse-grained \textbf{qubit-array mapper} decides the assignments of qubits to atom arrays. The two-qubit gates between atoms from two different arrays can be executed with atom movement, whereas intra-array two-qubit gates cannot be directly executed and require additional SWAPs. Therefore, the mapper tries to maximize the two-qubit gates between arrays by formulating the problem as a MAX k-cut (k is the number of arrays) of a constructed gate frequency graph and solving it with an efficient heuristic.

Subsequently, at a more fine-grained level, we propose a \textbf{qubit-atom mapper} that maps the qubits to specific atoms in the array. For the SLM array, the qubit-atom mapper considers the load balance among rows and columns to adhere to specific requirements, such as maintaining the order of rows and columns during movements and avoiding overlaps between two rows or two columns. Then, the qubit-atom mapper strategically maps AOD atoms with frequent two-qubit gates to consistent locations across different arrays to increase the execution parallelism.

After completion of this two-level mapping, our \textbf{high-parallelism AOD atom router} is activated. In one iteration, it identifies a front layer of non-dependent gates in the transpiled circuit and selects a subset of these gates that satisfies all the hardware constraints. Then, it performs the movements of AOD rows and columns and illuminates the lasers for gate execution. The router works iteratively until the whole quantum circuit is completed. 

We extensively evaluate \name on 20+ benchmarks of generic circuits (arbitrary, QASMBench, and SupermarQ circuits), Quantum simulation, and QAOA circuits with 5 to 100 qubits and $10$ to $10^4$ two-qubit gates.
We estimate the circuit fidelity under realistic parameters while comprehensively modeling four aspects of movement overhead, including heating, cooling overhead, decoherence, and atom loss.

On average, \name achieves 5.6\x, 3.4\x, 3.5\x, and 2.8\x in two-qubit gate reduction, 3.7\x, 3.5\x, 3.2\x, and 2.2\x in depth reduction over IBM Superconducting, FAA with long-range gates~\cite{baker2021exploiting}, FAA with rectangular topology, and FAA with triangular topology. \name is 1000\x faster in compilation than a solver-based~\cite{tan2022} compiler with similar fidelity. It also reduces the number of pulses by up to 6.5\x over Geyser~\cite{geyser}.

\begin{figure}[t]
    \centering
    \includegraphics[width=\columnwidth]{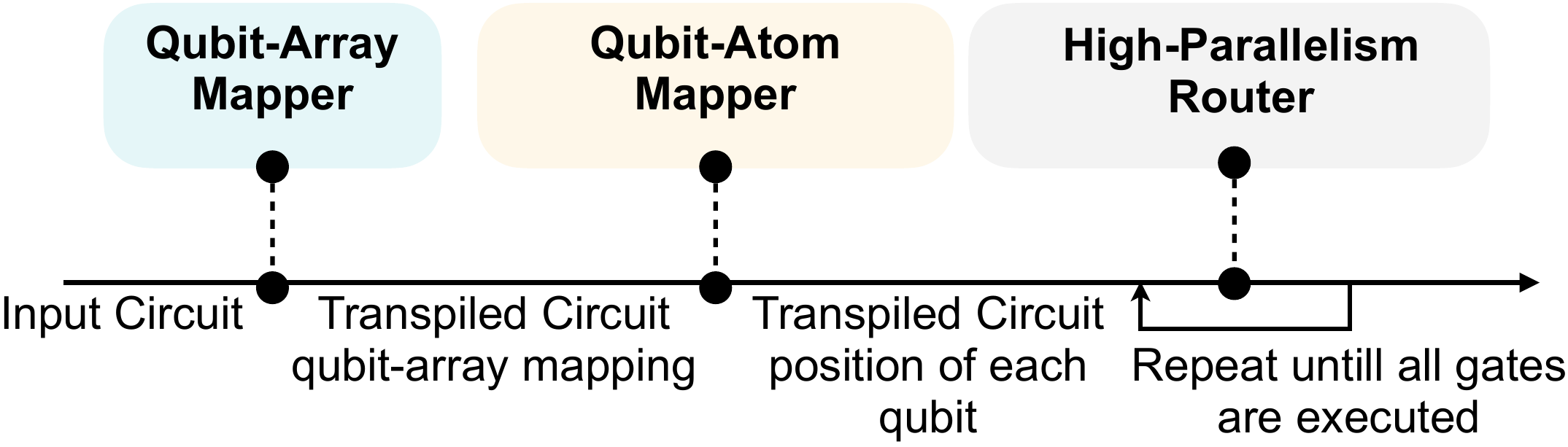}
    \caption{Overall pipeline of \name.}
    \label{fig:flow}
    \vspace{-15pt}
\end{figure}
\section{\raa Background}
\label{sec:background}

\noindent\textbf{\raa Qubit Movement.}
In \raas, atoms are held in two kinds of traps as illustrated in Fig.~\ref{fig:teaser}: fixed traps generated by a spatial light modulator (SLM) and movable traps created by crossed 2D acousto-optic deflectors (AODs). Each 2D AOD is composed of two intersecting 1D AODs, capturing atoms at the points of intersection. The system allows for manipulation of the X/Y coordinates of AOD columns/rows for aligned qubit movements. In particular, within the same AOD set, a column/row is restricted from crossing another (Fig.~\ref{fig:teaser} Challenge 2). However, columns/rows from different AOD sets can intersect. While the figure depicts two AOD sets, control over additional sets has been experimentally validated~\cite{graham_multi-qubit_2022}.

\noindent\textbf{\raa Gates.}
High-fidelity, individually addressable one-qubit gates are induced by a Raman laser (Fig.~\ref{fig:device_illustration} left)~\cite{bluvstein2022quantum, prl19-levine, graham_multi-qubit_2022}.
The two-qubit entangling gates are achieved through the Rydberg blockade mechanism~\cite{prl00-rydberg-blockade}.
When two atoms are within the \textit{Rydberg range} $r_b$ and illuminated by a Rydberg laser, a controlled $Z$-rotation is induced (Fig.~\ref{fig:device_illustration} right).
We need to move two qubits within $r_b$ of each other for two-qubit gates or to separate them by $\ge 2.5r_b$ when no gate is performed (Fig.~\ref{fig:teaser} Challenge 1).

\noindent
\noindent\textbf{Pros {\&} Cons of Neutral Atom Arrays in General.}
Neutral atoms enjoy the same benefits as other natural qubits, e.g., ions, compared to fabricated qubits, e.g., superconductors: high-fidelity gates, ease of calibration, and virtually limitless source.
Compared to electrical traps used by ions, optical traps for neutral atoms can be fabricated as 2D rather than 1D and thus scale to much larger sizes, e.g., 1020 SLM traps demonstrated in~\cite{ebadi2021quantum} which can, in theory, trap 1020 atoms and then later demonstrated trapping of more than 1200 atoms in\mbox{~\cite{norcia2024iterative}}.
In the long run, the scalability of \raas can still be challenging at a few-thousand-qubit scale because they are built from quantum optics experimental apparatus not initially designed for large-scale quantum computing.
However, this limitation is comparative, as other platforms lag behind in scaling so far.
Companies such as QuEra and PASQAL have begun to develop hardware with scalability in focus.

\noindent\textbf{Pros {\&} Cons of \raas vs FAA.} In FAA~\cite{graham_multi-qubit_2022, baker2021exploiting, geyser, Saffman_2016, beugnon2007two}, fixed SLM atoms restrict qubit connectivity, requiring an additional AOD array to modulate Rydberg interactions between selected adjacent pairs.
This limits scalability because twice as many SLM atoms and a similarly sized AOD array are needed compared to \raa. This also compromises $\cz$ fidelity; e.g., 92.5\% in FAA~\cite{graham_multi-qubit_2022} versus 97.5\% in \raa~\cite{bluvstein2022quantum}.
FAA needs SWAP gates ($\sim$3 $\cz$s) to route the qubits for two-qubit gates.
In contrast, \raa's AOD movements are high-fidelity, limited primarily by coherence time. Ref.~\cite{bluvstein2022quantum} estimates that with only $0.1\%$ of coherence time lost, an AOD array could traverse a region hosting $\sim2,000$ qubits.
A drawback of atom movement and transfer is the risk of atom loss, jeopardizing the integrity of the entire quantum circuit.
Thus, a careful comparison of these two routing methods is required, as we shall see in this paper.

\noindent
\noindent\textbf{Application-Specific Compilation.}
We contrast \name with three existing application-specific compilers: ZZ~\cite{alam2020circuit,alam2020efficient,alam2020noise}, 2QAN~\cite{lao20222qan}, and Paulihedral~\cite{li2022paulihedral}. ZZ focuses on the commutation of ZZ gates in QAOA; our approach generalizes this by avoiding suboptimal layering of ZZ gates. 2QAN minimizes SWAPs by absorbing them into existing two-qubit unitaries, a technique less relevant for \raas, which instead employs atom movement for routing. Paulihedral presents two strategies: 1) Optimizing Pauli term order to reduce circuit depth, a technique orthogonal to our work and applicable to pre-routing; 2) Tailoring each Pauli term for limited-connectivity NISQ hardware, but less applicable to \raas.

\section{\name Compilation Framework}
\label{sec:qubit-mapping-routing}

\begin{figure}[t]
    \centering
    \includegraphics[width=\columnwidth]{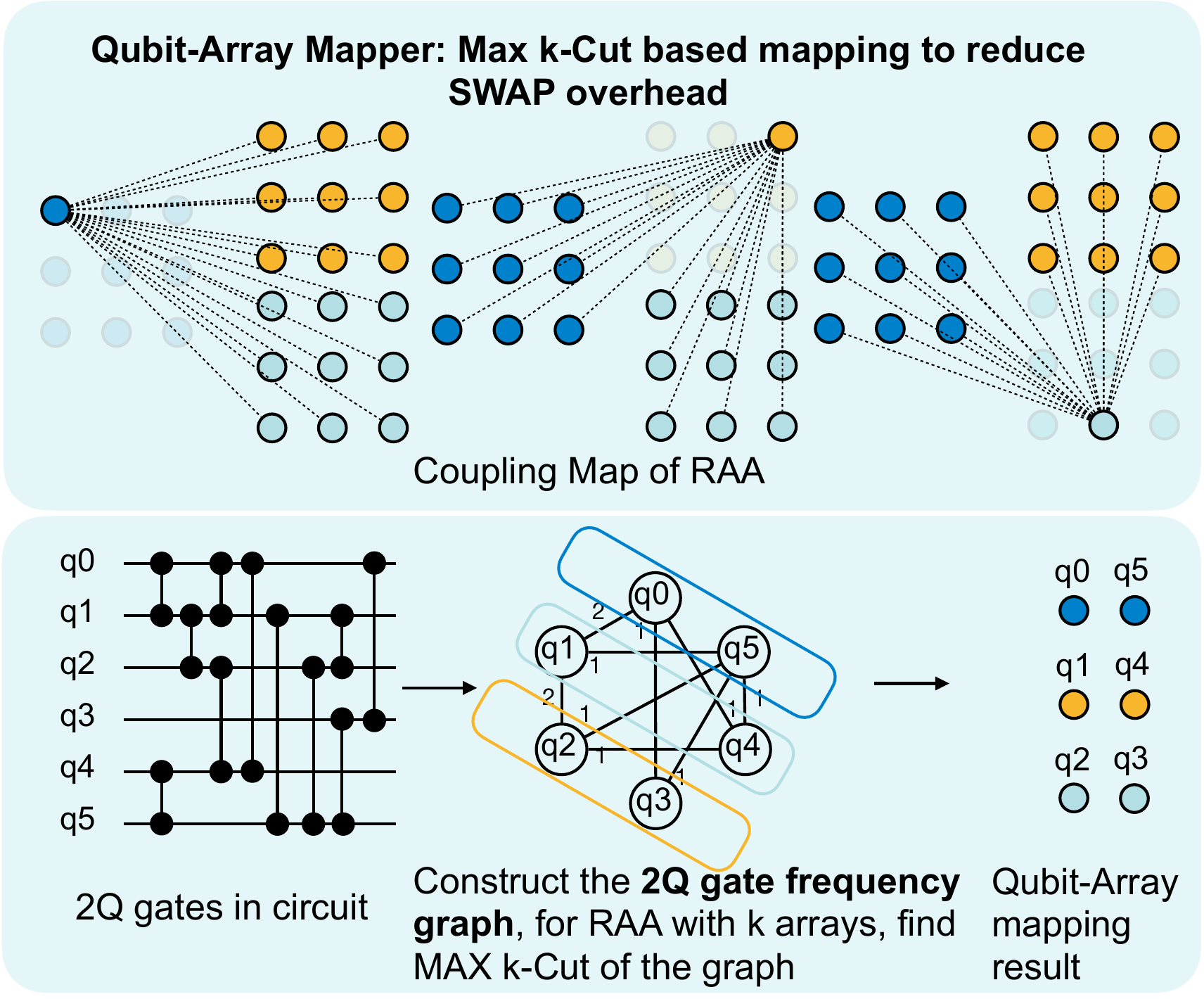}
    \vspace{-8pt}
    \caption{Qubit-array mapper decides the array for each qubit with MAX k-Cut to reduce SWAP overhead.}
    \label{fig:qubit_array_mapper}
    \vspace{-10pt}
\end{figure}

The overall compilation flow is shown in Fig.~\ref{fig:flow}. Our compiler contains a \textit{qubit-array mapper} to determine the array in which each qubit should be placed and a \textit{ qubit atom mapper} to decide the specific position of each qubit inside the mapped array. After deciding all qubit positions, we also design a \textit{high-parallelism router} to generate the movements and gate schedules of the rows and columns of the AOD array, respecting hardware constraints and maximizing parallelism.

As shown in Fig.~\ref{fig:qubit_array_mapper}, the \raa features one SLM array and multiple AOD arrays, represented by different colors in the Fig.~\ref{fig:qubit_array_mapper} top. Qubits are fixed in the SLM array, and atom pairs in the SLM array will never fall within the Rydberg radius, precluding intra-array two-qubit gates. For atom pairs in the same AOD array, our compiler also avoids intra-array two-qubit gates because of the risk of atom loss.
Consequently, two-qubit gates can only be performed between two different arrays. Since there is only one SLM array, at least one of the two arrays will be an AOD array, and thus the gate can be executed via atom movement. This can be represented by a complete multipartite graph for feasible two-qubit gate operations. As a result, the atoms in the same array are equivalent from the coupling map perspective. 
Therefore, we tackle the mapping problem hierarchically: firstly, deciding which array a qubit should be mapped to, and secondly, determining the specific position for the qubit in that array.
\subsection{Qubit-Array Mapper}

\begin{algorithm}[t]
\footnotesize{
    \caption{Qubit-Array Mapper}
    \label{alg:alg_heuristic_router}
    \SetAlgoLined
    \SetKwInOut{Input}{Input}
    \SetKwInOut{Output}{Output}

    \Input{Quantum circuit $C$ with $n$ qubits}
    \Input{Number of AODs $m$}
    \Output{Mapping of qubits to AODs in dictionary $M$}

    $E \gets n \times n$ zero matrix; \tcp{Adjacency matrix for the gate frequency graph.}
    $M \gets$ empty dictionary; \tcp{Maps qubits to AODs, M[i] are the qubits that are mapped to ith AOD}

    \For{each 2Q gate $G$ in $C$}{
        $E[i][j]+=\gamma^l$; \tcp{$G$ acts on qubit $i$ and $j$, and lies in the $l$-th layer of the circuit.}
    }

    \For{$i = 1$ to $n$}{
        $Bestcut \gets 0$; \\
        $Mapto \gets 0$; \\
        \For{$j = 1$ to $m$}{
            $Currentcut \gets \sum_{k \notin M[j]} E[i][k]$; \\
            \If{$Bestcut < Currentcut$}{
                $Bestcut \gets Currentcut$; \\
                $Mapto \gets j$; \\
            }
        }
        Add $i$ to $M[Mapto]$;
    }
}

\end{algorithm}

\label{sec:QubitArrayMapper}
\begin{figure}[t]
    \centering
    \includegraphics[width=1\columnwidth]{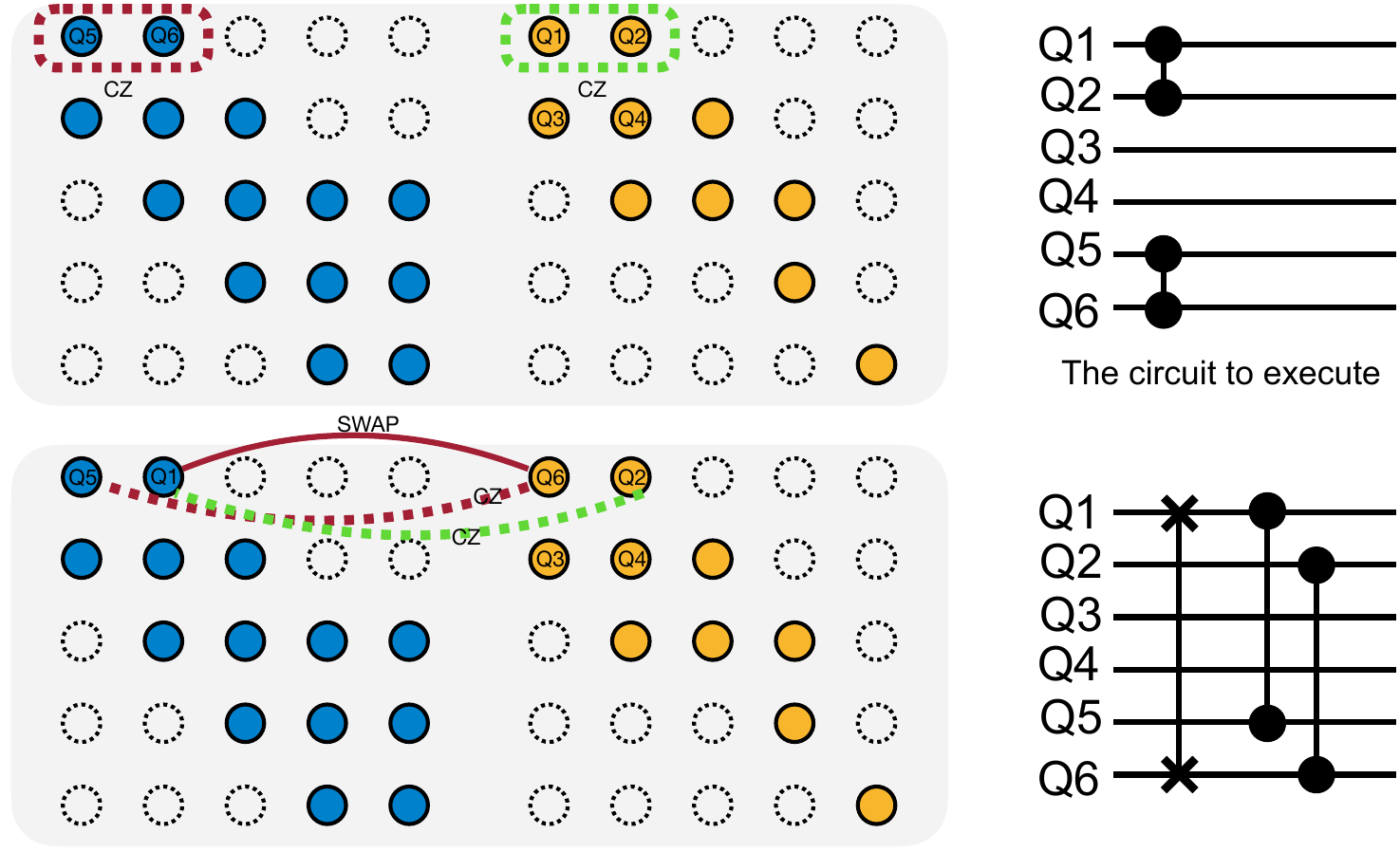}
    \vspace{-16pt}
    \caption{An example showing why SWAP is still needed and how they are applied.}
    \label{fig:swap_example}
    \vspace{-15pt}
\end{figure}

Ideally, if all gates in a circuit involve qubits in different arrays, there will be no SWAP overhead.
To this end, \name tries to find a mapping to maximize the inter-array gates, and thus minimize SWAP overhead. 
This goal can be approached with a \textit{two-qubit gate frequency graph} as shown in Fig.~\ref{fig:qubit_array_mapper} bottom. In the graph, each vertex represents a qubit and an edge indicates that a gate exists between two qubits. The edge weights are determined by the frequency of 2-qubit gates between the qubit pairs. More gates between a qubit pair will result in a large edge weight. Then, finding the mapping that allows the most directly executable inter-array two-qubit gates is translated to a \textbf{MAX k-Cut} of the gate frequency graph, where the graph is divided into k partitions, corresponding to k arrays in \raa, with the aim of maximizing the summation of edge weights crossing different partitions. Although this problem is NP-hard, we adopt a straightforward greedy algorithm that yields an approximation of $1-\frac{1}{k}$~\cite{coja2003max}.
The greedy algorithm decides the partition assignment of each vertex one by one while ensuring that each vertex can maximize the cut between partitions of already assigned vertices. The algorithm is outlined in Alg.~\ref{alg:alg_heuristic_router}.

\begin{figure}[t]
    \centering
    \includegraphics[width=\columnwidth]{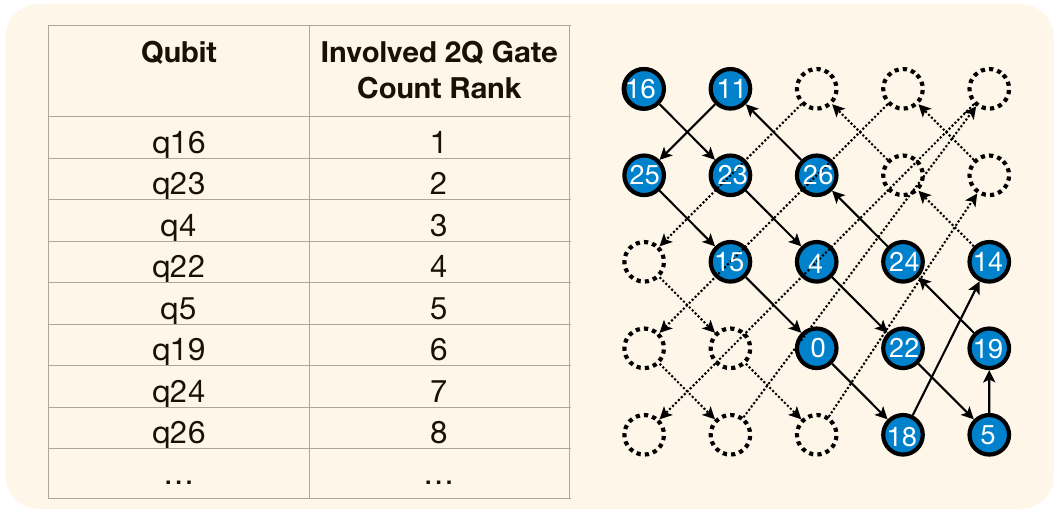}
    \vspace{-15pt}
    \caption{Qubit-atom mapper first maps SLM qubits with ``load balance mapping" to avoid potential constraint violations.}
    \label{fig:qubit_atom_mapper0}
    \vspace{-10pt}
\end{figure}

\begin{figure}[t]
    \centering
    \includegraphics[width=\columnwidth]{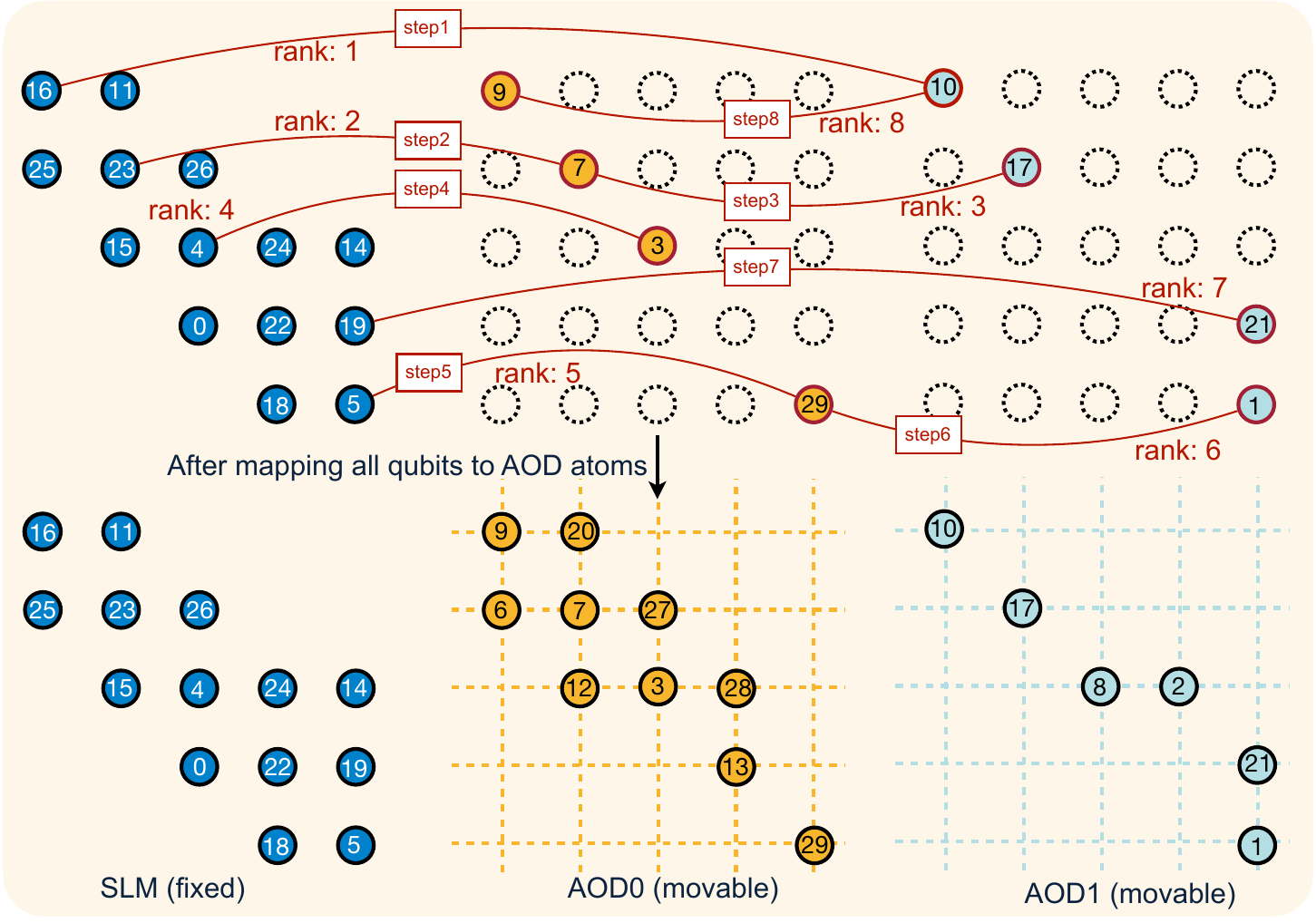}
    \vspace{-5pt}
    \caption{Qubit-atom mapper then maps AOD qubits with "aligned AOD Mapping" based on gate frequency rank. 
    }
    \label{fig:qubit_atom_mapper1}
    \vspace{-10pt}
\end{figure}

For the calculation of the specific edge weight, each gate between a qubit pair will contribute a weight that exponentially decreases according to how many layers the gate is from the first layer in the circuit DAG and is controlled by a factor $\gamma$. We add decay because, for gates in the later layers, we have less control over the qubit positions, and the benefit this gate can have from a better mapping is less.

After finding the initial qubit-array mappings, we leverage the default SABRE~\cite{li2019tackling} in Qiskit with the multipartite coupling graph to insert SWAP gates and pass the mapping and transpiled circuit to the qubit-atom mapper.
Fig.\mbox{~\ref{fig:swap_example}} provides an example of how SWAP is inserted. In this instance, our aim is to execute two-qubit gates between Q1 and Q2, as well as between Q5 and Q6. However, due to hardware limitations, neither operation is directly feasible; certain movements may lead to undesired interactions, while others are physically prohibited. We address this issue by introducing an additional SWAP insertion step. Specifically, we swap Q1 with Q6 using three \mbox{\cz\xspace} and one-qubit gates, enabling the direct execution of all the two-qubit gates.

\subsection{Qubit-Atom Mapper}
\begin{figure}[t]
    \centering
    \includegraphics[width=\columnwidth]{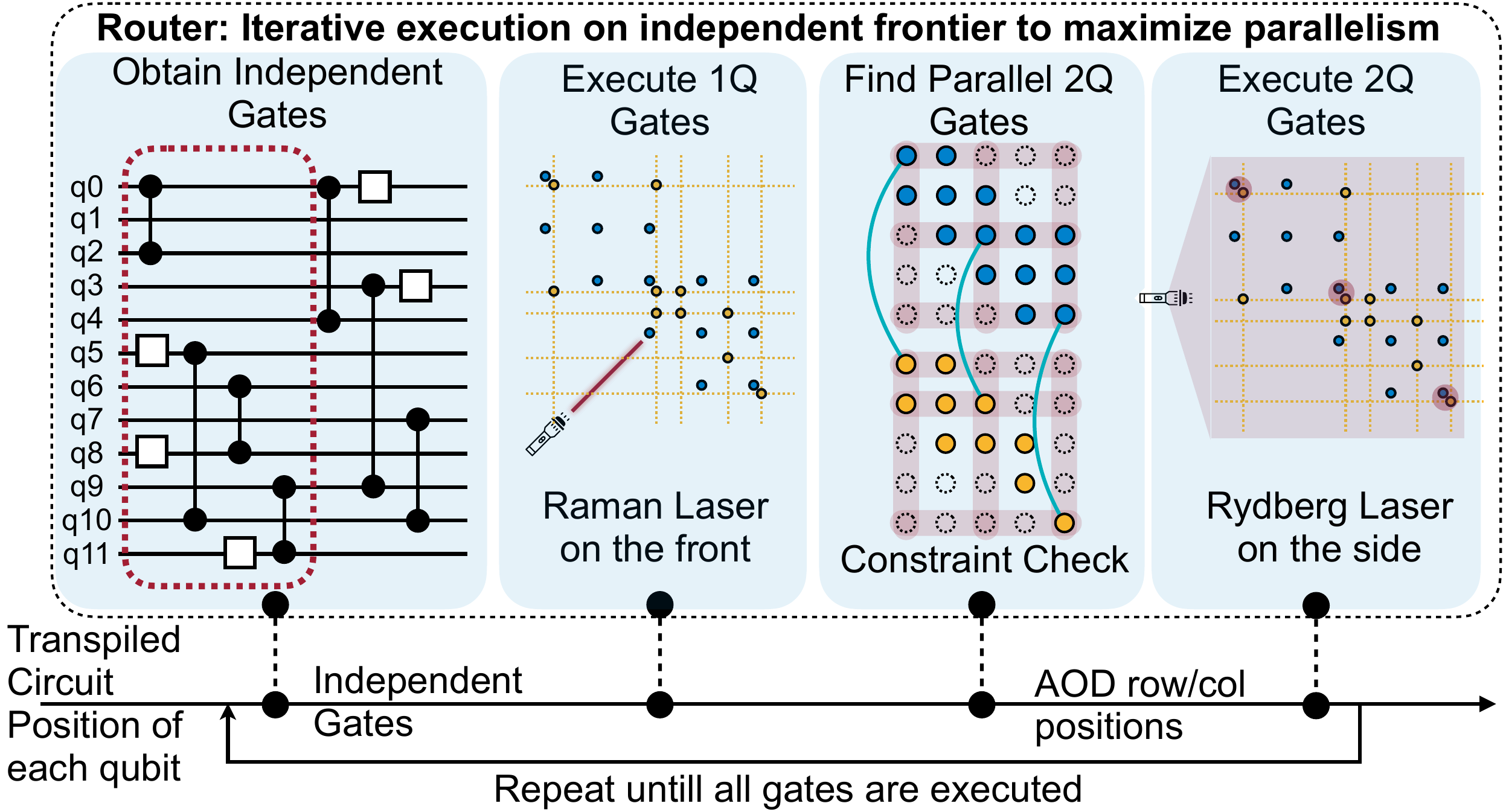}
    \vspace{-5pt}
    \caption{The pipeline of the high-parallelism router.}
    \label{fig:router}
    \vspace{-10pt}
\end{figure}
\label{sec:QubitAtomMapper}
\name further decides the fine-grained position of each qubit in each array. This step does not impact the SWAP overhead. Instead, it essentially focuses on selecting the optimal atom positions in the array to minimize circuit depth. It is crucial to note that once the atoms are positioned, the absolute positions of the SLM atoms and the relative positions of AOD atoms are immutable during circuit execution. Suboptimal mapping can diminish opportunities for parallel gate execution, thus elongating the circuit depth. In the worst-case scenario, all two-qubit gates need to be executed sequentially, even without circuit-level dependencies. Therefore, the primary goal of our qubit-to-atom mapping is to maximize parallel-gate executions. To achieve this, we perform qubit-atom mapping in two steps: firstly, map all SLM qubits, and secondly, map all AOD qubits. 

\begin{figure}[t]
    \centering
    \includegraphics[width=\columnwidth]{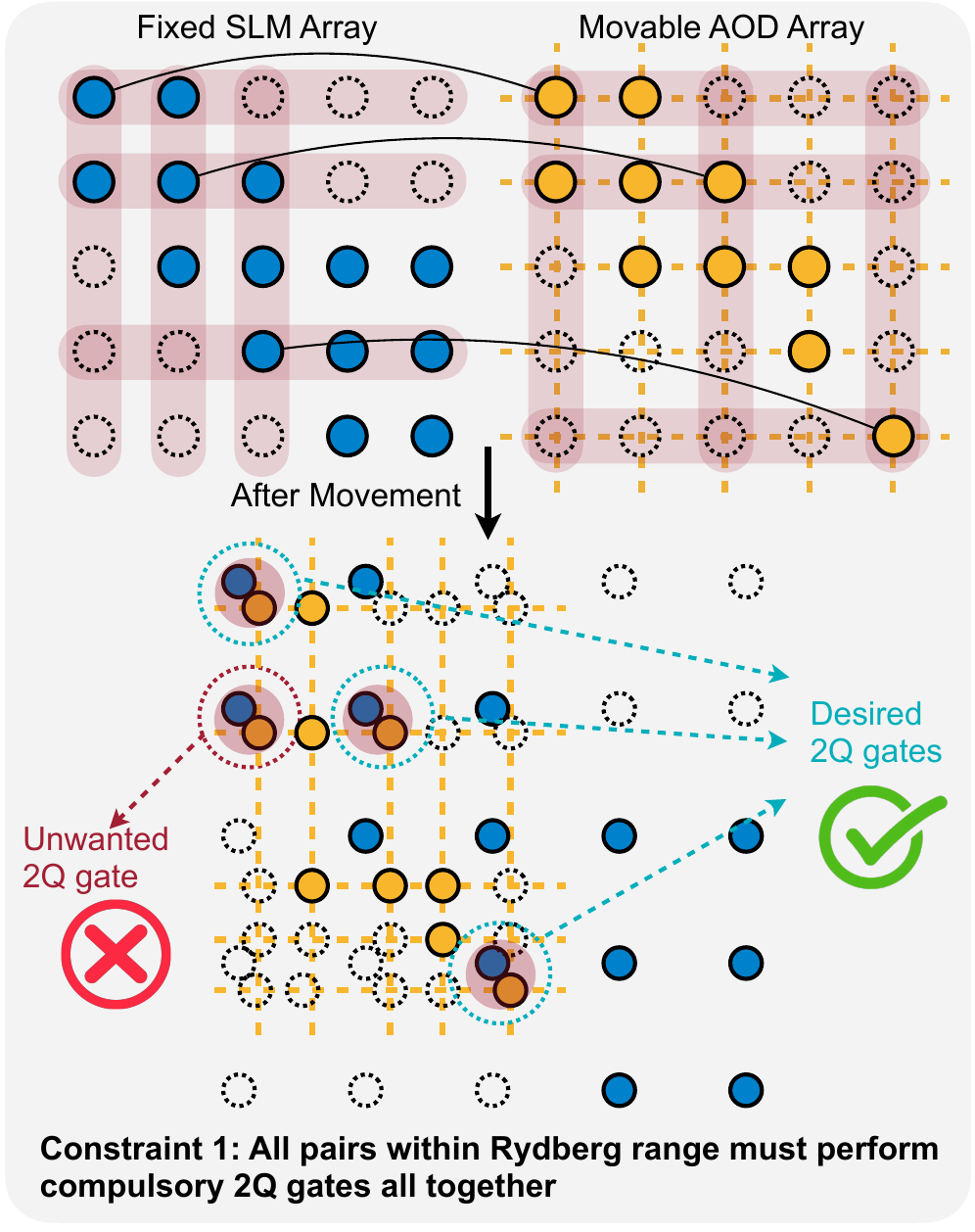}
    \caption{The first constraint checked by the router.}
    \label{fig:router_checks}
    \vspace{-20pt}
\end{figure}

First, it is crucial for qubit-atom mapping of SLM qubits following the design principle of load balance between rows/columns. Specifically, we need to balance the number of atoms across rows because most qubits concentrating in a few rows will create more conflicts within the same row. The column load balance is similar.
To this end, we propose the \textbf{load balance SLM mapping}, as illustrated in Fig.~\ref{fig:qubit_atom_mapper0}.
We first sort the qubits in descending order based on the number of two-qubit gates they involved. Then, we map the qubits to atoms with a topological order starting from the upper-left corner, prioritizing filling the diagonal first and subsequently forming a spiral trajectory. This will ensure that the row-wise/column-wise sum of involved two-qubit gates of qubits is balanced across all rows/columns. The load balance can, in return, maximally avoid violations of the constraint1 on unwanted gates (Fig.~\ref{fig:router_checks}) and constraint3 that rows/columns cannot overlap as illustrated (Fig.~\ref{fig:router_checks2}).
Two atoms from a near-diagonal region have a greater chance of differing in row/column index.

\begin{figure}[t]
    \centering
    \includegraphics[width=\columnwidth]{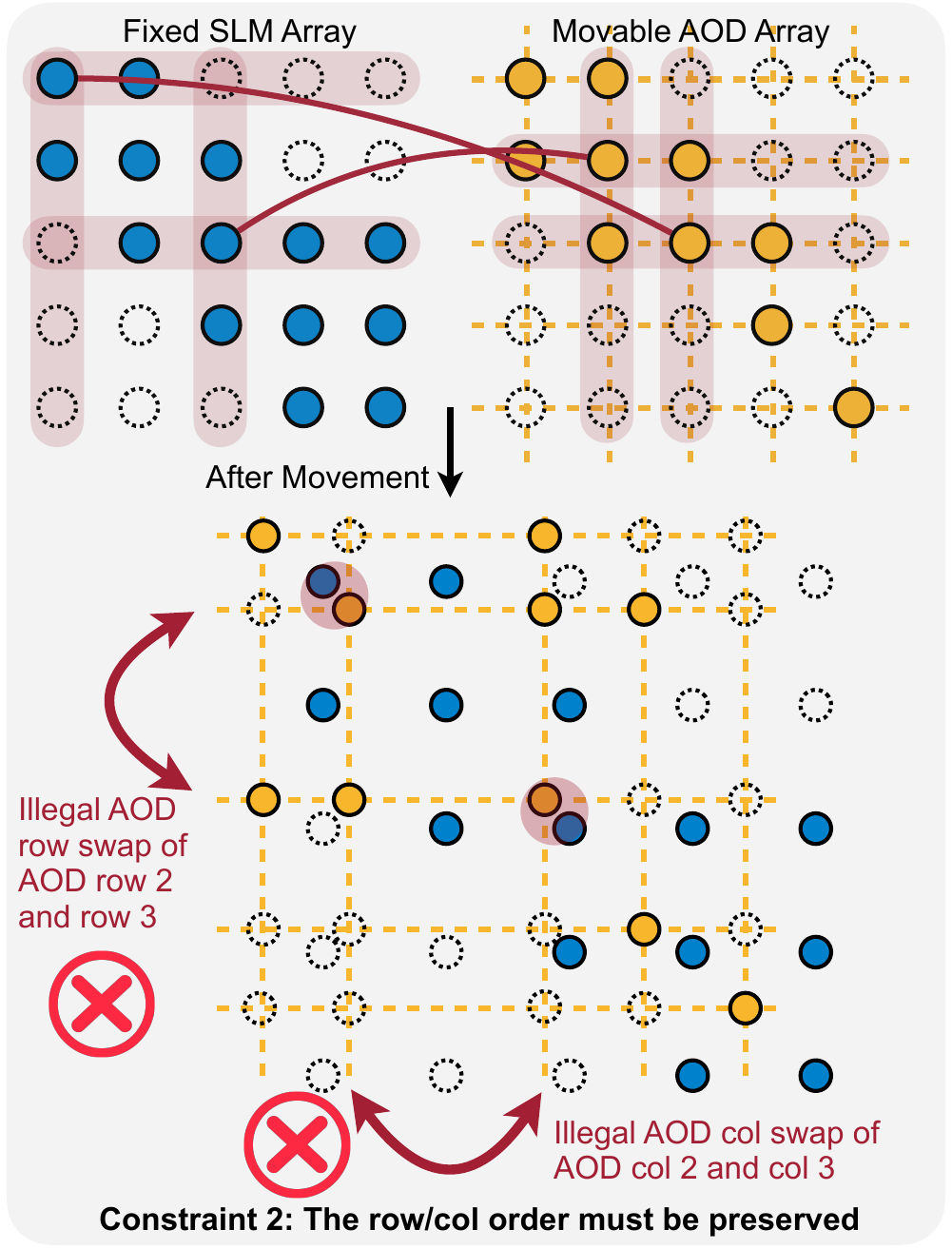}
    \vspace{-5pt}
    \caption{The second constraint checked by the router.}
    \label{fig:router_checks1}
    \vspace{-15pt}
\end{figure}

Second, we need to map the AOD qubits. The design principle for AOD mapping is position alignment for frequent qubit pairs. Specifically, we propose \textbf{aligned AOD mapping} to make sure that the pairs of atoms of the same positions in two arrays will have highly frequency two-qubit gates so that we naturally enable higher parallelism if we move and align two arrays. We first sort the frequency of qubit pairs with two-qubit gates; then, the frequent pairs will be mapped to the same spatial location in two arrays as the example shown in Fig.~\ref{fig:qubit_atom_mapper1}. The SLM qubits atom mapping has already been determined. The ``rank" is determined by the two-qubit gate frequency in decreasing order. The Q16 and Q10 have the highest frequency, so we map Q10 to the same top left corner as Q16 in the SLM array. Similarly, for the second rank gate between Q23 and Q7, we map Q7 to the same row2 column2 position in the yellow AOD0 as Q23 in SLM.
This mapping strategy will avoid the violation of Constraint 2 on row / column orders (Fig.~\ref{fig:router_checks1}) because the high-frequency gates will have a small relative displacement.

\subsection{High-Parallelism AOD Router}
\label{sec:AODRouter}
With the specific qubit-atom mappings and the transpiled circuit, we propose a router for AOD atom movements and gate scheduling. The router's responsibility is to generate the instructions on the movement of rows and columns of all AODs and decide when to turn on the lasers to perform gates. Meanwhile, it must respect all the hardware constraints during movements and optimize for high parallelism.

The pipeline is depicted in Fig.~\ref{fig:router}. The router first identifies independent ``frontier gates'' from a direct acyclic graph (DAG) circuit representation. Then, it executes all single-qubit gates by illuminating the Raman lasers. Although two-qubit gates do not have circuit-level dependencies and could theoretically be executed in parallel, moving them all to the corresponding atom may violate the hardware constraints. Therefore, the router greedily finds maximally legal parallel two-qubit gates. Starting with a single gate, the router incrementally adds gates to the set, assessing three hardware constraints. If violated, the two-qubit gate will not be added to the set but will be pushed back to the circuit DAG. After finding the legal parallel two-qubit gate set, the router will move the AOD rows and columns to the target positions and turn on the Rydberg laser to perform two-qubit gates. This router runs iteratively until the entire circuit is finished.

\begin{figure}[t]
    \centering
    \includegraphics[width=\columnwidth]{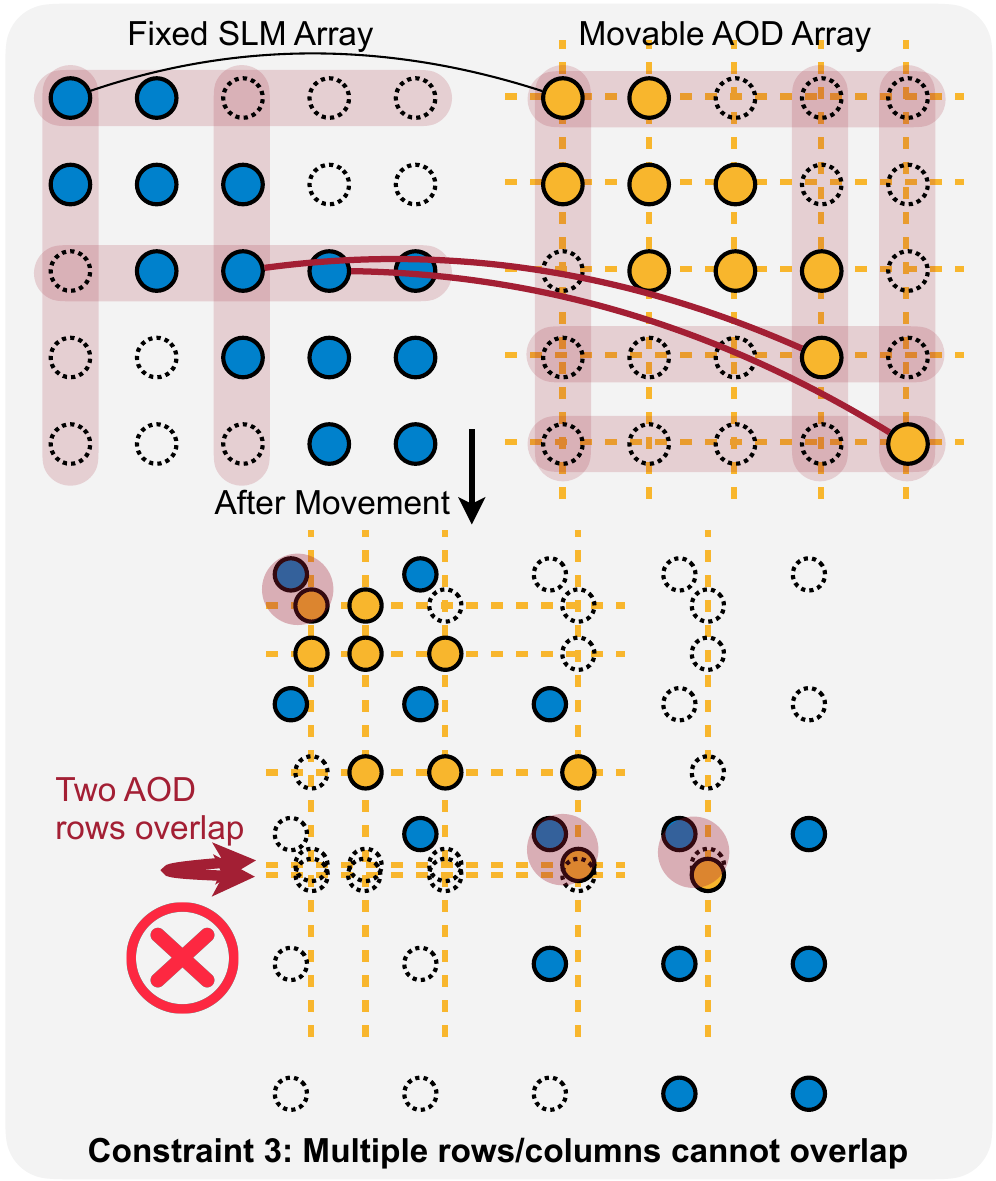}
    \vspace{-5pt}
    \caption{The third constraint checked by the router.}
    \label{fig:router_checks2}
    \vspace{-15pt}
\end{figure}

The first hardware constraint is that all pairs within the Rydberg range must perform gates together as in Fig.~\ref{fig:router_checks}. In the figure's top part, three black connection lines indicate the valid gates in the circuit. If we perform the movement to perform those gates simultaneously, as in the bottom part, we will introduce an unwanted gate marked in red. Therefore, this set of gates cannot be executed in parallel. The second hardware constraint is shown in Fig.~\ref{fig:router_checks1}: the row order and column order of AOD arrays must be preserved, i.e., there cannot be a position swap of two rows/columns. In the figure, we show that to execute two two-qubit gates annotated by two red lines on the top, we have to swap row 2 and row 3 and swap column 2 and column 3, as illustrated in the bottom part. This is illegal and thus will not be accepted by the router. The third constraint is that the AOD rows and columns cannot overlap as in Fig.~\ref{fig:router_checks2}. To finish the two gate pairs connected by red lines, AOD rows 4 and 5 must be moved to the same position and overlap, which is not allowed. 

\section{Atom Movement Overhead}
\label{sec:movement}

We characterize atomic motion overhead through four metrics, as formulated in Eq.~{\ref{eq:fmov}}: additional two-qubit error due to heating, $F_{\text{mov\_heating}}$; atom loss due to heating, $F_{\text{mov\_loss}}$; additional overhead due to cooling, $F_{\text{mov\_cooling}}$; and decoherence during movement, $F_{\text{mov\_deco}}$.
\begin{equation}\label{eq:fmov}
F_{\text{mov}} = F_{\text{mov\_heating}} \times F_{\text{mov\_loss}} \times F_{\text{mov\_cooling}} \times F_{\text{mov\_deco}} 
\end{equation}
\noindent\textbf{Atom Heating.} Atom movement causes heating and degrades two-qubit gate fidelity, necessitating the computation of the ``vibrational quantum number'', $n_{\text{vib}}$, for each atom. This quantity encapsulates an atom's cumulative vibrational energy due to heating. Our framework tracks each atom's $n_{\text{vib}}$ based on motion history.
\begin{equation}
\label{eq:fheat}
    F_{\text{mov\_heating}} = \prod_{i=1}^{N_{\text{2Q}}} \left(1 - \lambda \times (1-f_{\text{2Q}}) \times n_{\text{vib}}(i)\right)
\end{equation}
According to {\cite{bluvstein2022quantum}}, heating's fidelity impact is proportional to both two-qubit gate infidelity, ($1-f_{\text{2Q}}$), and $n_{\text{vib}}$, as described in Eq.~{\ref{eq:fheat}}. In interactions involving an AOD and an SLM atom, the $n_{\text{vib}}$ of the AOD atom is considered; for AOD-AOD interactions, the $n_{\text{vib}}$ values are summed.
\begin{figure}[h]
    \centering
    \includegraphics[width=\columnwidth]{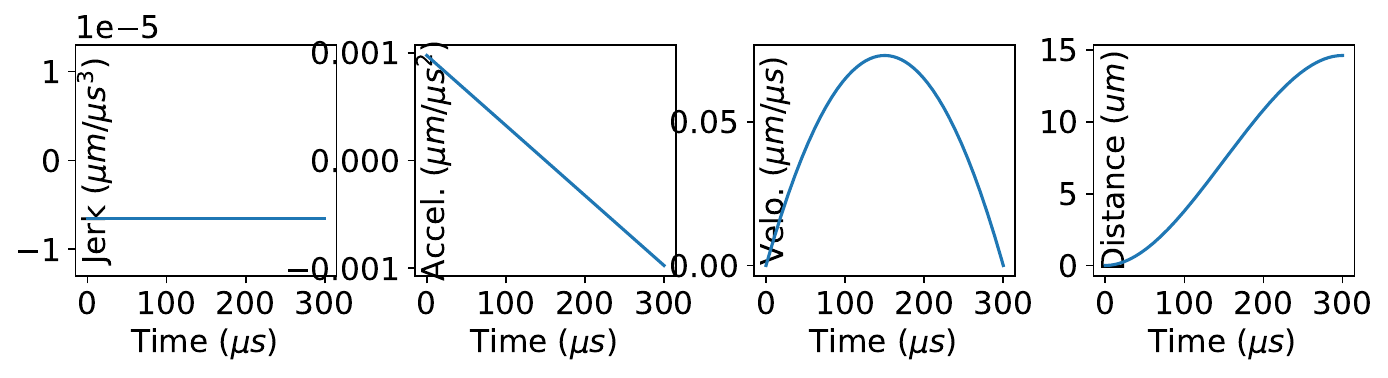}
    \vspace{-20pt}
    \caption{The atom movement pattern.}
    \label{fig:move_pattern}
    \vspace{-10pt}
\end{figure}

Ref.~\cite{bluvstein2022quantum} propose
a constant negative jerk (the derivative of acceleration) strategy as delineated in Figure~{\ref{fig:move_pattern}} to minimize $n_{\text{vib}}$, resulting in a linearly decreasing acceleration and a parabolic velocity profile. The incremental change in $n_{\text{vib}}$ for a single movement is:
$
\Delta n_{\text{vib}}=\frac{1}{2}\left(\frac{6D/x_{\text{zpf}}}{\omega_0^2 T_{\text{mov}}^2}\right)^2
$,
where $D$ is the movement distance; $x_{\text{zpf}} \equiv \sqrt{h/2\pi \cdot (2 m \omega_0)}$ is the zero-point size of the particle; $h$ is the Planck‘s constant; $\omega_0$ is the trap frequency and $T_{\text{mov}}$ is the duration of this move.
From this equation, we can obtain the insights: $n_{\text{vib}} \propto N_{\text{move}} / T_{\text{mov}}^4$, indicating that a minor increase in $T_{\text{mov}}$ allows a substantially greater $N_{\text{move}}$ before the heating error is too large, albeit at the cost of increased decoherence error. This trade-off is further analyzed in Section~{\ref{sec:evaluation}}.

Here, we analyze when doing atom movement is better than adding SWAPs. According to~{\cite{bluvstein2022quantum}}, $x_{\text{zpf}} = 38$nm, and $\omega_0$ can be between 40KHz and 80KHz. We use 80KHz. Movement time $T_{mov}$ can be between 200$\upmu$s to 400$\upmu$s and we select 300$\upmu$s. Atom distance, the separation between the rows and columns of the atom, needs to be greater than 6 times the Rydberg radius, which is $6\times2.5=15\upmu m$. According to~{\cite{bluvstein2022quantum}} Extended Fig.~5d movement path and Extended Fig.~6c in which the finite-temperature (Doppler) error rate is 0.3\%, we compute the $\lambda$ in Eq.~{\ref{eq:fheat}} as $0.109$.
With these parameters, movement-induced $n_{\text{vib}}$ increases are calculated to be $\Delta n_{\text{vib}} = 0.0054$ for a 1 hop movement, $0.13$ for 5 hops, and $0.54$ for 10 hops. In the experiments shown in Sec.~{\ref{sec:evaluation}}, the cost of routing two-qubit gates is around 2$\times$ the logical two-qubit gates in FAA. Eq.~{\ref{eq:fheat}} says the additional error incurred by heating is $\lambda n_{\text{vib}}$ times the two-qubit error. So, when $n_{\text{vib}} <\frac{2}{\lambda}\approx 18$, which corresponds to fewer than 1000 gates, we expect better performance using \raa atom movement, since the overhead of movement will be less than that of additional SWAP gates.

\noindent\textbf{Atom Loss.} The heating and vibration also increase the atom loss probability. According to~{\cite{bluvstein2022quantum}}, the atom loss probability of moving and atom loss fidelity are:
 $   P_{\text{mov\_loss}} = 1 - \frac{1}{2} \left(1 + \text{erf} \left[ \frac{n_{\text{vib}}^{\text{max}}-n_{\text{vib}}}{\sqrt{2 n_{\text{vib}}}}\right]\right)$,
$    F_{\text{mov\_loss}} = \prod_{i=1}^{N_{\text{move}}}\prod_{j} (1-P_{\text{mov\_loss}}(i, j)),$
where $j$ indicates the moved atoms in a movement. In~{\cite{bluvstein2022quantum}}, $n_{\text{vib}}^{\text{max}}=33$.
When $n_{\text{vib}}=30$, $F_{\text{mov\_loss}}=0.708$; when $n_{\text{vib}}=20$, $F_{\text{mov\_loss}}=0.998$; when $n_{\text{vib}}=15$, $F_{\text{mov\_loss}}=0.999998$. We see a fast improvement as $n_{\text{vib}}$ is reduced. Thus, we establish a $n_{\text{vib}}$ threshold of 15 for cooling, beyond which the impact on atom loss is negligible.

\begin{table}[t]
\centering
\renewcommand*{\arraystretch}{1}
\setlength{\tabcolsep}{0.5pt}

\caption{Hardware Parameters}

\resizebox{\columnwidth}{!}{%
\begin{tabular}{l|c|c|c|c|c|c}
\toprule
Parameter &2Q fidelity & 1Q fidelity& 2Q gate T&1Q gate T& Coherence T&Atom distance\\
\midrule
Neu. Atom & 0.9975\cite{bluvstein2022quantum} & 0.99992\cite{bluvstein2022quantum} & 380ns\cite{bluvstein2022quantum} & 625ns\cite{bluvstein2022quantum} &15s\cite{bluvstein2022quantum} &15$\upmu$m\cite{bluvstein2022quantum}\\
Supercon. & 0.9975 &0.99992 &480ns\cite{ibmq} & 35.2ns\cite{ibmq}& 801.2$\upmu$s\cite{ibmq}&-\\
\midrule
\midrule
Parameter &T per move & Atom transfer T&Atom loss P&$x_{\text{zpf}}$&$\omega_0$&$\lambda$\\
\midrule
Neu. Atom & 
300$\upmu$s\cite{bluvstein2022quantum}  & $15\upmu$s\cite{bluvstein2022quantum} & 0.0068\cite{covey20192000} & 38nm\cite{bluvstein2022quantum} & 80kHz\cite{bluvstein2022quantum} & 0.109\cite{bluvstein2022quantum}\\
\bottomrule
\end{tabular}
}

\label{tab:hyperparams}
\vspace{-20pt}

\end{table}

\noindent\textbf{Cooling Overhead.} When any atom's $n_{\text{vib}}$ exceeds this threshold, cooling is initiated for the entire AOD array by swapping its quantum information with a pre-prepared, cooled AOD array initialized to the zero state. This information transfer requires two CZ gates for each atom, so the overhead is below. When the threshold is 15, cooling is required around every 100 gates.
    $F_{\text{mov\_cooling}} = \prod_{i}^{N_\text{cooling}} f_\text{2Q}^{2\times N_\text{AOD}}.$

\noindent\textbf{Decoherence.} Finally, we model the decoherence due to the time overhead of moving atoms. 
$F_{\text{mov\_deco}} =\prod_i \exp(-N_i\times T_{\text{mov,i}}/T_1)$,
where $T_\text{mov,i}$ is the moving time of stage $i$, and $N_i$ is the number of qubits, including AOD and SLM used at stage $i$. 
In assessing fidelity overheads, we juxtapose additional SWAPs in FAA with decoherence costs in \raa. According to~\cite{bluvstein2022quantum}, the fidelity for a two-qubit (two-qubit) gate is $f_{\text{2Q}}=0.975$, and $T_1=1.5s$. In the next section, our experiments show that the routing overhead for each two-qubit gate in FAA is around two additional two-qubit gates, reducing the overall fidelity by $f_{\text{2Q}}^2=0.95$.
For \raa, the decoherence overhead of one additional two-qubit gate in a 10-qubit circuit is $F_{\text{mov\_deco}} = \exp(-300 \times 10^{-6}/1.5 \times 10)=0.998$. This overhead scales to $0.99$ and $0.98$ for 50- and 100-qubit circuits, respectively. Consequently, \raa's overhead is smaller than adding two two-qubit gates in FAA.
Remarkably, this advantage will persist in the forecasted future. Suppose that both $f_{\text{2Q}}$ and $T_1$ are scaled by an order of magnitude to $0.9975$ and $15s$, respectively. The revised overhead for 10, 50, and 100 qubits becomes $0.9998$, $0.999$, $0.998$, while the FAA routing overhead will be $0.995$, maintaining \raa's benefit.

\section{Evaluation}\label{sec:evaluation}
\begin{figure*}[t]
    \centering
    \includegraphics[width=\textwidth]{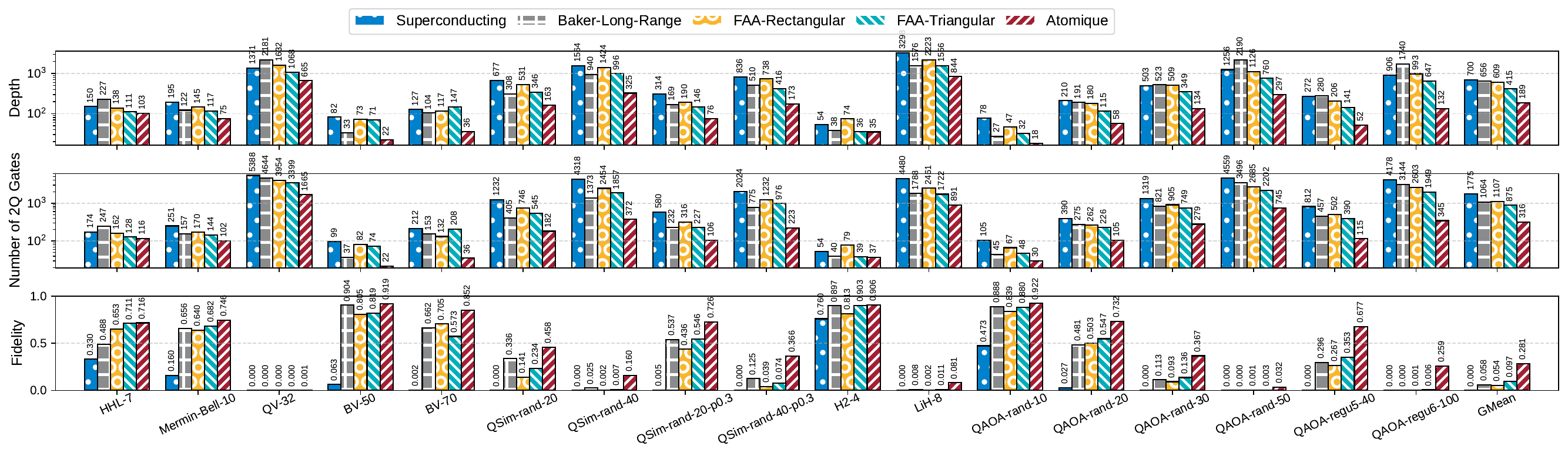}
    \vspace{-15pt}
    \caption{Circuit depth, two-qubit gate count, and fidelity comparisons between different architectures.}
    \label{fig:maintable_fid_n2q_depth}
    \vspace{-10pt}
\end{figure*}

\begin{table}[t]
\centering
\renewcommand*{\arraystretch}{1}
\setlength{\tabcolsep}{1pt}

\caption{Benchmarks}

\resizebox{\columnwidth}{!}{%
\begin{tabular}{l|ccccccc|cc}
\toprule
Name & Type & Dataset & Num. & Num. & Num. & 2Q Gate & Degree & Tan-Solver & Tan-IterP\\
 &  &  & Qubits & 2Q gates & 1Q gates & per Q & per Q & & \\
\midrule
HHL & Generic & QASMBench & 7 & 196 & 794 & 56.0 & 5.7 & {\color{red}timeout} & {\color{green}solved}\\
Mermin-Bell & Generic & SupermarQ & 10 & 67 & 30 & 13.4 & 7.6 &  {\color{red}timeout} & {\color{green}solved}\\

QV & Generic & QASMBench & 32 & 1536 & 4096 & 96 & 19.7 &  {\color{red}timeout} & {\color{green}solved}\\

BV50 & Generic & QASMBench & 50 & 22 & 100 & 0.9 & 0.9 &  {\color{red}timeout} & {\color{green}solved}\\

BV70 & Generic & QASMBench & 70 & 36 & 417 & 1.0 & 1.0 &  {\color{red}timeout} & {\color{green}solved}\\

\midrule
QSim-rand-20 & QSim & - & 20 & 180 & 275 & 18.0 & 4.2 &  {\color{red}timeout} & {\color{green}solved}\\
QSim-rand-40 & QSim & - & 40 & 380 & 567 & 19.0 & 5.7 &  {\color{red}timeout} & {\color{green}solved}\\
H2 & QSim & Molecule & 4 & 40 & 54 & 20.0 & 3.0 & {\color{green}solved} & {\color{green}solved}\\
LiH & QSim & Molecule & 6 & 1134 & 1602 & 283.5 & 6.0 &  {\color{red}timeout} & {\color{green}solved}\\
\midrule
QAOA-rand-10 & QAOA & - & 10 & 27 & 10 & 5.4 & 5.4 &  {\color{red}timeout} & {\color{green}solved}\\
QAOA-rand-20 & QAOA & - & 20 & 80 & 20 & 8.0 & 8.0 &  {\color{red}timeout} & {\color{green}solved}\\
QAOA-regu5-40 & QAOA & - & 40 & 100 & 40 & 5.0 & 5.0 &  {\color{red}timeout} & {\color{green}solved}\\
QAOA-regu6-100 & QAOA & - & 100 & 300 & 100 & 6.0 & 6.0 &  {\color{red}timeout} &  {\color{red}timeout}\\
\midrule
\midrule
Mermin-Bell & Generic & SupermarQ & 5 & 19 & 15 & 7.6 & 4.0  & {\color{green}solved} & {\color{green}solved}\\
VQE & Generic & SupermarQ & 10 & 9 & 40 & 1.8 & 1.8 & {\color{green}solved} & {\color{green}solved}\\
VQE & Generic & SupermarQ & 20 & 19 & 80 & 1.9 & 1.9 & {\color{green}solved} & {\color{green}solved}\\
Adder & Generic & QASMBench & 10 & 65 & 101 & 13.0 & 2.6 & {\color{green}solved}  & {\color{green}solved}\\
BV & Generic & QASMBench & 14 & 13 & 81 & 1.9 & 1.9 & {\color{green}solved} & {\color{green}solved}\\
\midrule
QSim-rand-5 & QSim & - & 5 & 20 & 29 & 8.0 & 2.4 &  {\color{green}solved} & {\color{green}solved}\\
QSim-rand-10 & QSim & - & 10 & 80 & 122 & 16.0 & 4.6 & {\color{green}solved} & {\color{green}solved}\\
\midrule
QAOA-rand-5 & QAOA & - & 5 & 6 & 5 & 2.4 & 2.4 & {\color{green}solved} & {\color{green}solved}\\
QAOA-regu3-20 & QAOA & - & 20 & 30 & 20 & 3.0 & 3.0 & {\color{green}solved} & {\color{green}solved}\\
QAOA-regu4-10 & QAOA & - & 10 & 20 & 10 & 4.0 & 4.0 & {\color{green}solved}& {\color{green}solved}\\
\bottomrule
\end{tabular}
}

\label{tab:new_benchmarks}
\vspace{-10pt}
\end{table}

\subsection{Evaluation Methodology}
\noindent\textbf{Fidelity Estimation.}
We model the fidelity with four terms: $F = F_{\text{1Q}} \times F_{\text{2Q}} \times  F_{\text{transfer}} \times F_{\text{mov}}.$ $F_\text{mov}$ has been discussed in Sec.~\ref{sec:movement}. $F_{\text{1Q}}$ is the fidelity impact of executing one-qubit gates:
$F_{\text{1Q}} = f_{\text{1Q}}^{N_{\text{1Q}}} \times \exp(-T_{\text{1Q}}/T_1 \times N)$,
where $f_{\text{1Q}}$ is the fidelity of a single one-qubit gate; $N_{\text{1Q}}$ is the total number of one-qubit gate after compilation; $T_{\text{1Q}}$ is the cumulative time of one-qubit gates; $T_1$ is the coherence time; and N is the total number of qubits. Two-qubit fidelity is similar:
$F_{\text{2Q}} = f_{\text{2Q}}^{N_{\text{2Q}}} \times \exp(-T_{\text{2Q}}/T_1 \times N)$.
$F_{\text{transfer}}$ is the fidelity impact of performing atom transfer between SLM and AOD:
$F_{\text{transfer}} = (1-P_{\text{loss}})^{N_{\text{transfer}}} \times \exp(-T_{\text{transfer}}/T_1 \times N)$,
where $P_{\text{loss}}$ is the atom loss probability; $N_{\text{transfer}}$ is the number of transfers; and $T_{\text{transfer}}$ is the cumulative time of transfers.

We build our own fidelity model because current commercial hardware or simulators cannot satisfy our requirements. For example, the hardware available from QuEra is fundamentally different from ours because it lacks movable atoms for now. Also, their hardware supports only pulse-level analog quantum computing, and the corresponding simulator, Bloqade, supports only Hamiltonian simulation. Neither supports the application of a quantum gate, requiring us to program the Hamiltonian directly. Investigating the correct pulse for a gate is beyond the scope of this paper. Finally, Bloqade does not support noisy simulation, thus preventing us from evaluating the performance of various compilers under noise.

\begin{figure}[t]
    \centering
    \includegraphics[width=1\columnwidth]{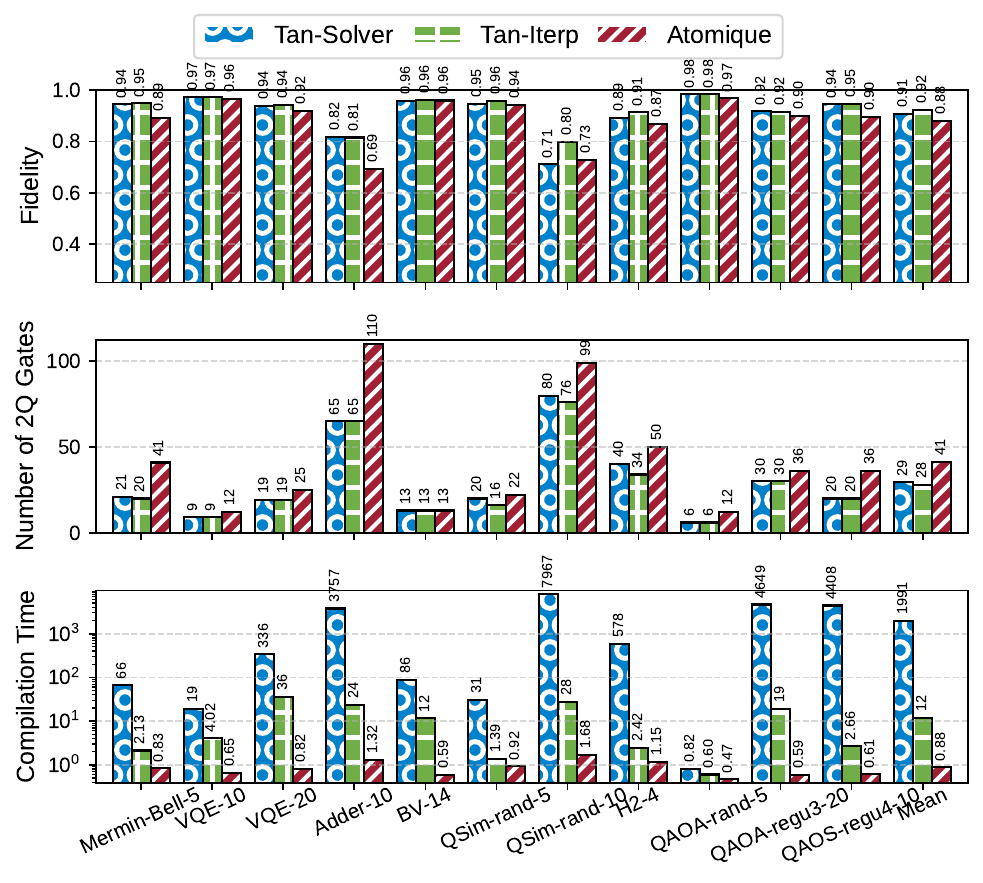}
    \vspace{-15pt}
    \caption{Comparisons to Tan-Solver and Tan-IterP. \name has similar fidelity but over 1000\x faster.}
    \label{fig:solver_comp_bar}
    \vspace{-15pt}

\end{figure}
\noindent\textbf{Hardware Parameters.} In the neutral atom device cited from \mbox{\cite{bluvstein2022quantum}}, a CZ gate is realized through two global Rydberg pulses, each lasting 190ns, leading to a two-qubit gate duration of 380ns and fidelity of 0.9975. Arbitrary single-qubit gates employ a `BB1' pulse sequence, achieving a one-qubit gate time of $625$ns and a fidelity of 0.9992, calculated based on a scattering error of $2 \times 10^{-4}$ per $\pi$ pulse. The atom transfer time is $15\upmu s$ with a loss probability of 0.0068~\mbox{\cite{baker2021exploiting}}. 
Parameters such as SLM atom distance, move time, coherence time, $x_{\text{zpf}}, \omega_0, \lambda$ have been addressed in Section~\mbox{\ref{sec:movement}}. 
Our default configuration is 10$\times$10 topology with 1 SLM array and 2 AOD arrays. For baseline architectures, we equalize qubit numbers with those reported in \name.
For superconducting qubits, parameters are derived from the IBMQ Experience platform~\cite{ibmq}. Fidelity metrics for two-qubit and one-qubit gates align with those for neutral atoms for unbiased comparisons. We scale up the coherence time for superconducting and neutral atom devices by 10x and scale down their two-qubit and one-qubit gate errors to make evaluation on large quantum circuits possible. Tab.~\ref{tab:hyperparams} summarizes the parameters.


\noindent\textbf{Benchmarks.}
To assess compiler performance, we utilize three benchmark categories: algorithmic (generic), quantum simulation (QSim), and Quantum Approximate Optimization Algorithm (QAOA). Their characteristics are detailed in Table~\mbox{\ref{tab:new_benchmarks}}. The metric ``Degree per Q'' represents the average count of unique qubits that interact with a given qubit.
Algorithmic circuits are sourced from SupermarQ~\cite{tomesh2022supermarq} and QASMBench~\cite{10.1145/3550488}. QSim circuits are randomly generated with a probability of 0.5 for a qubit to exhibit a non-$I$ Pauli operator, and each circuit comprises ten Pauli strings. For molecular Hamiltonians, we consider H2 and LiH.
QAOA circuits are constructed by randomly placing $\zz$ gates between all pairs of qubits with a probability of 0.5. In addition, QAOA circuits based on regular graphs are generated, in which $\zz$ interactions are placed to qubit pairs with an edge in the regular graph. These benchmarks span a broad spectrum featuring a variety number of qubits and circuit depth.

\begin{figure}[t!]
    \centering
    \includegraphics[width=\columnwidth]{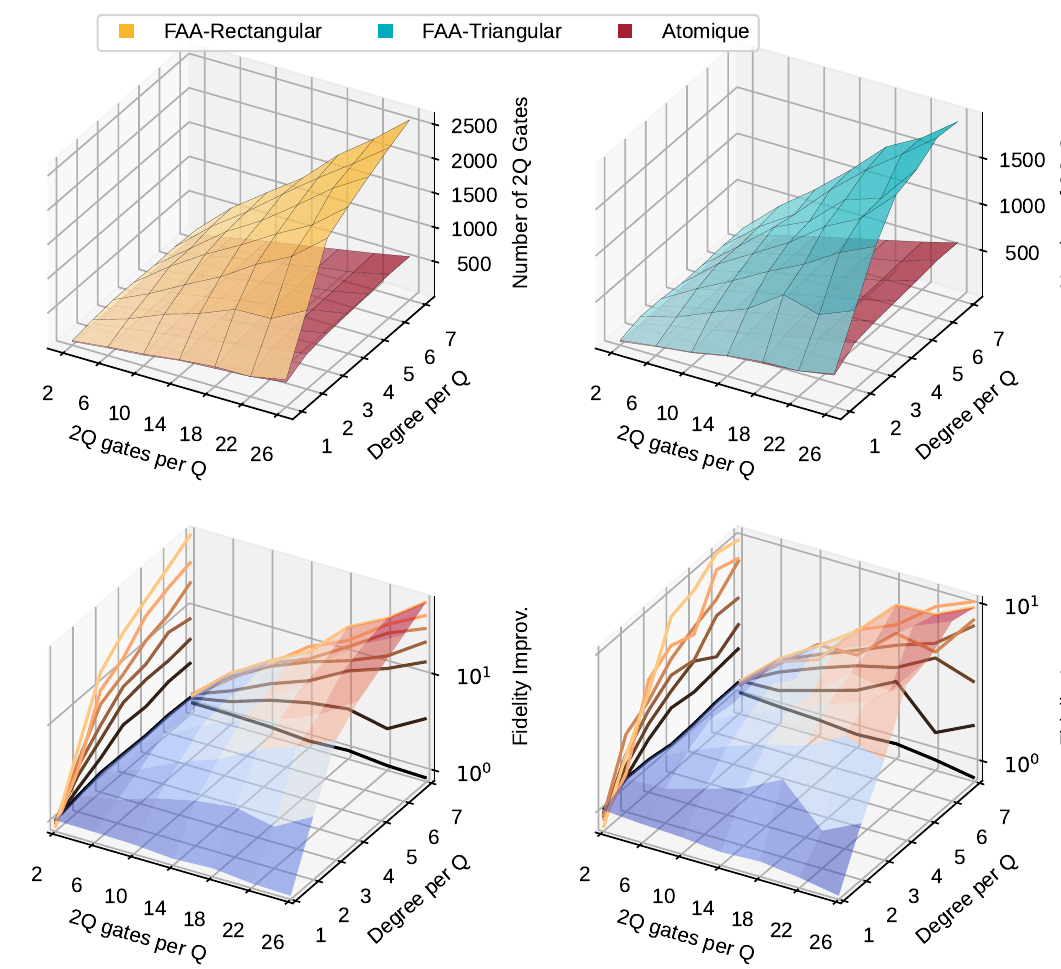}
    \vspace{-20pt}
    \caption{Performance of generic circuits with different characteristics on different architectures.}
    \label{fig:3d_n2q_fid_arb}
\end{figure}

\begin{figure}[t]
    \centering
    \includegraphics[width=\columnwidth]{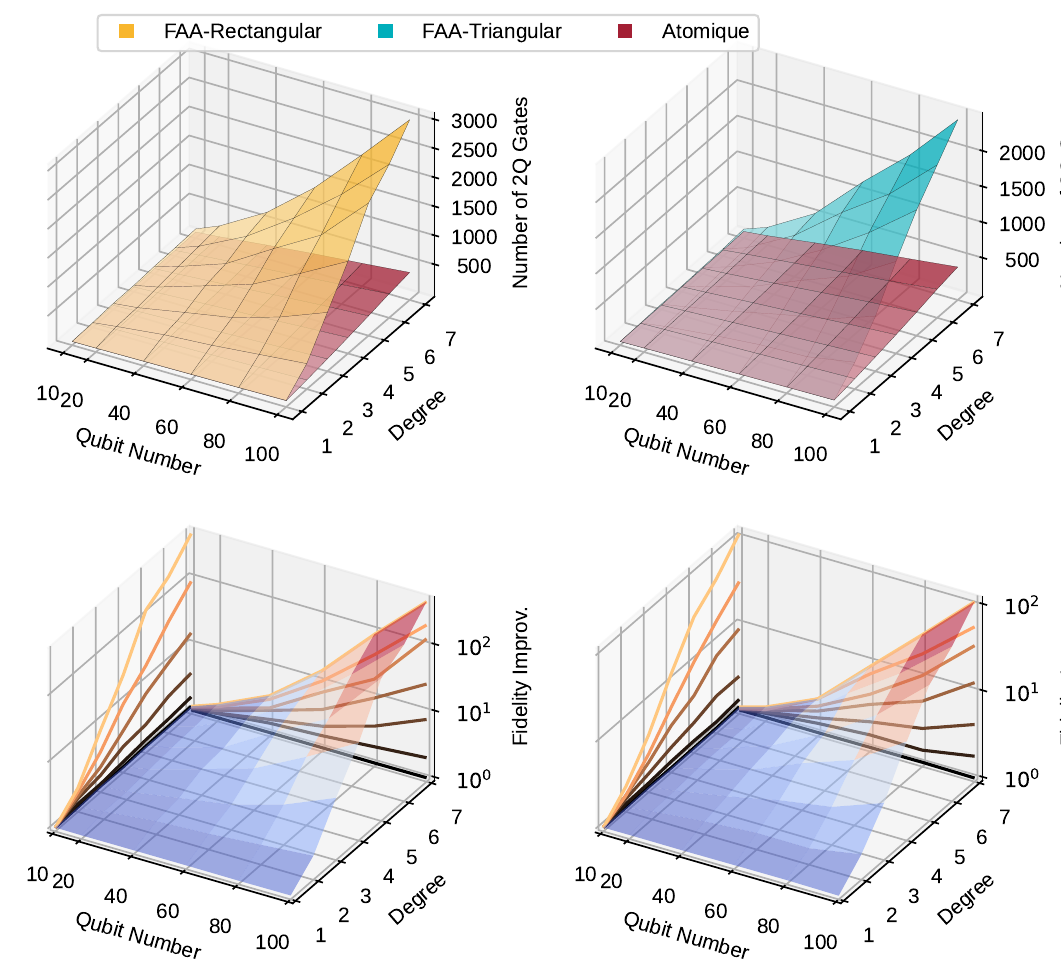}
    \vspace{-20pt}
    \caption{Performance of QAOA circuits with different characteristics on different architectures.}
    \label{fig:3d_n2q_fid_qaoa}
    \vspace{-15pt}
\end{figure}

\begin{figure}[t]
    \centering
    \includegraphics[width=\columnwidth]{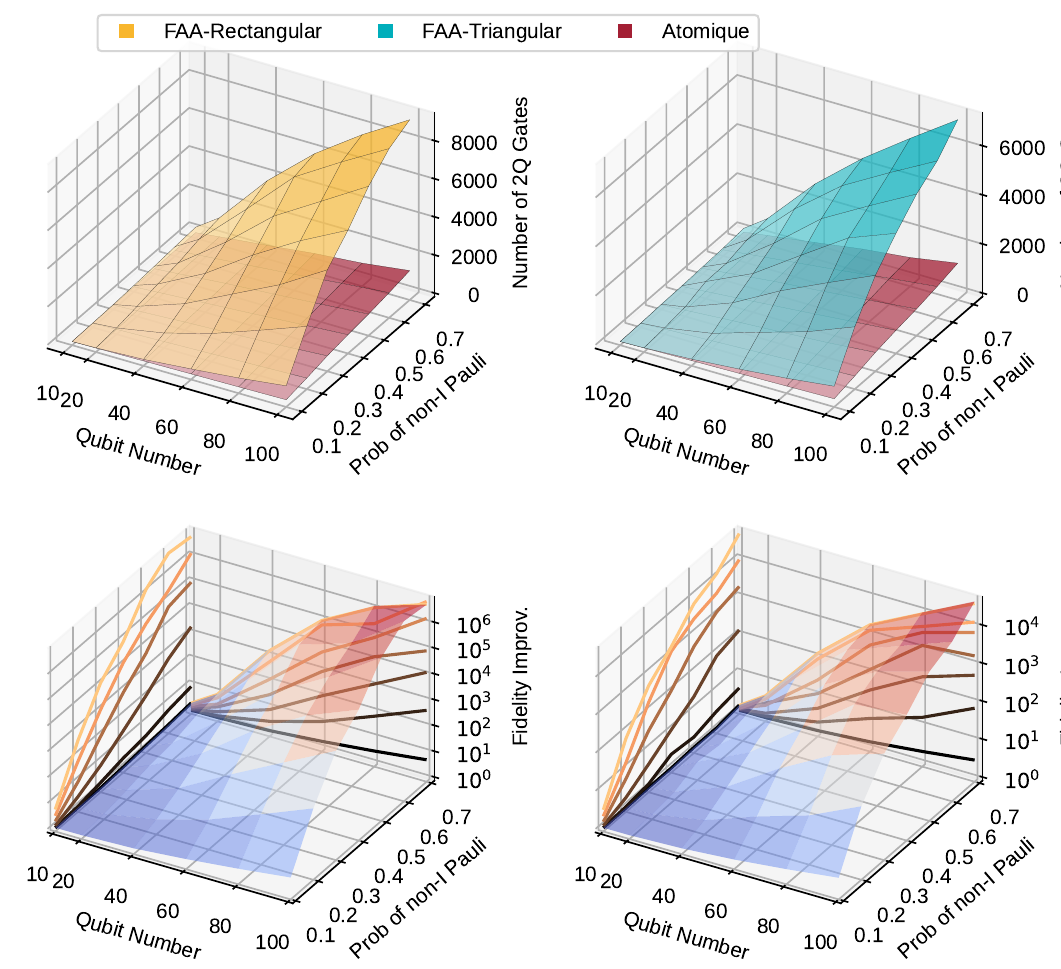}
    \vspace{-20pt}
    \caption{Performance of Quantum Simulation circuits with different characteristics on different architectures.}
    \label{fig:3d_n2q_fid_qsim}
    \vspace{-10pt}
\end{figure}

\noindent\textbf{Baselines.} To show the unique advantage of \raa, we establish a comparison with four other distinct architectures, each using a specialized compiler, along with two additional compilers for \raa.
(1) \textbf{Superconducting}: Utilizing IBM's 127-qubit Washington superconducting machine with a heavy hexagon coupling graph.
(2) \textbf{Baker-Long-Range}: A fixed rectangular array of atoms with long-range interactions. The compiler is as proposed by Baker et al.~\cite{baker2021exploiting}, and the maximum interaction range is set to four times the Rydberg radius. We use the \href{https://github.com/AndrewLitteken/neutral-atom-compilation}{open-source} by the authors.
(3) \textbf{FAA-Rectangular}: A fixed rectangular array of atoms that permits only nearest-neighbor interactions.
(4) \textbf{FAA-Triangular}: Fixed triangular arrays of atoms introduced in~\cite{geyser}. 
(5) \textbf{Tan-Solver} and \textbf{Tan-IterP}: we compare our compiler with two solver-based compilers~\cite{tan2022, tan2023compiling}.
Ref.~\cite{tan2023compiling} provides a tool, OLSQ-DPQA, where a parameter `optimal ratio' can be set to switch between the two methods: 1 for \textbf{Tan-Solver} and to 0 for \textbf{Tan-IterP}.
In the latter case, the SMT formulation is relaxed in a greedy manner, `iterative peeling', to support compiling larger circuits.
A few settings are worth mentioning: 1) we let OLSQ-DPQA use 16x16 atoms per SLM or AOD array, 2) the internal Z3 SMT solver~\cite{tacas08-demoura-bjorner-z3-smt-solver} has version 4.13.0.0, and 3) we impose a 24-hour timeout.
For experimental consistency, all tests run on a server equipped with a 40-core Intel(R) Xeon(R) Silver 4114 CPU @ 2.20GHz and 192GB of RAM. All baselines are using Qiskit Optimization Level 3 with SABRE~\cite{li2019tackling}.

\subsection{Main Results}
\noindent\textbf{Comparison of \raa with Other Quantum Architectures.}
Fig.~\ref{fig:maintable_fid_n2q_depth} shows \name outperforming four alternative architectures in fidelity, two-qubit gate count, and depth (number of parallel two-qubit layers), exceeding the strongest baseline, FAA-Triangular, by factors of 2.7$\times$, 2.8$\times$, and 2.2$\times$, respectively. Large and complicated circuits such as QSim-rand benefit the most due to higher connectivity and lower SWAP overhead. In simpler circuits, such as H2 simulations, different architectures perform comparably.

\begin{figure*}[t]
    \centering
    \includegraphics[width=\textwidth]{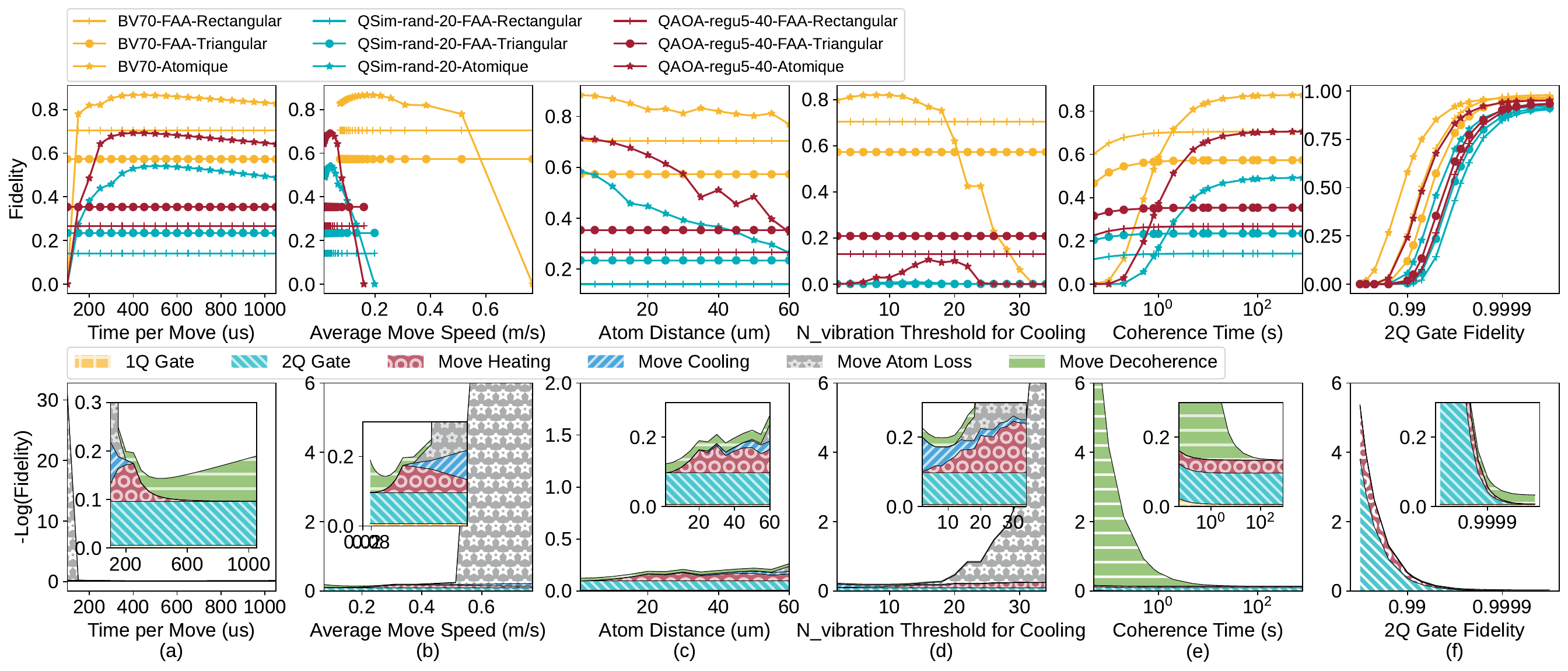}
    \vspace{-15pt}
    \caption{Sensitivity analysis of various hardware parameters.}
    \label{fig:ablation_2d_all}
    \vspace{-16pt}
\end{figure*}

\begin{figure}[ht]
    \centering
    \includegraphics[width=0.9\columnwidth]{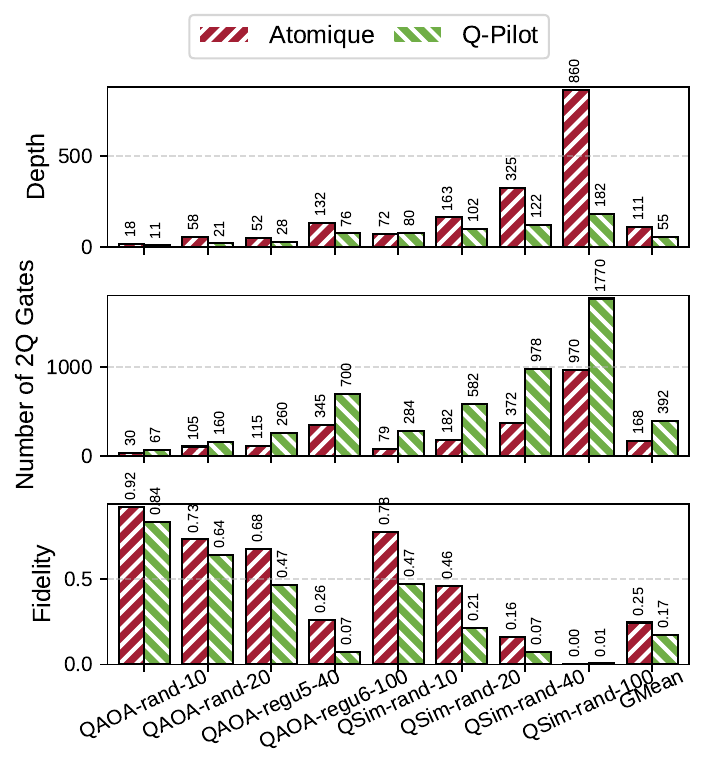}
    \vspace{-15pt}
    \caption{Comparison of \name and Q-Pilot.}
    \label{fig:flying}
    \vspace{-20pt}
\end{figure}

\noindent\textbf{Comparison of \name with Other Compilers on \raa.}
As indicated in the last column of Table~\mbox{\ref{tab:new_benchmarks}}, Tan-Solver~\cite{tan2022} is infeasible for circuits with more than 20 qubits within a 24-hour limit. Consequently, smaller circuits are used for comparisons in Fig.~\mbox{\ref{fig:solver_comp_bar}}. For a fair comparison, \mbox{\name} employs a single AOD, as two baselines lack multi-AOD support. \mbox{\name} maintains an average fidelity of 88\%, comparable to solver-based compilers, while being more than 2,000 times faster for small-scale problems. This speed advantage is exacerbated for larger problems as the solver-based compiler times out due to exponential time and memory requirements. 

We also compare \name with a recently-proposed compiler Q-Pilot~\cite{wang2023q}, which focuses on reducing circuit depth for specific applications (QAOA and QSIM). Fig.~\ref{fig:flying} shows that although Q-Pilot has less depth than \name, \name has better overall fidelity, showing a better balance between depth, two-qubit gates, and atom movement.

\noindent\textbf{Comparison of \name with Geyser.} Geyser~\cite{geyser} suggests using multi-qubit gates to synthesize quantum circuit targets by minimizing the pulses used to perform multi-qubit gates. Tab.~\ref{tab:comp_geyser} shows the number of pulses used for multi-qubit gates for our compiler and Geyser. 
Since we lack experimental data to demonstrate the fidelity of multi-qubit gates, we follow Geyser's recommendation by employing the number of pulses as a proxy for the gate's fidelity. The more pulses a gate uses, the less fidelity it has. A gate involving $n$ qubits necessitates $2n-1$ pulses. We limit the maximum function call in the dual-annealing process to $10^5$ to ensure equitable comparisons. The data presented in the table clearly indicates that our compiler significantly surpasses Geyser, achieving a reduction of up to \textbf{6.5$\times$} in the number of pulses.

\subsection{Analysis}

\noindent\textbf{What Kind of Circuits are More Suitable for \mbox{\raa}?}

\textbf{(a) Generic (Arbitrary) Circuits.} We analyze the efficacy of \mbox{\raa} on generic circuits in Fig.~\mbox{\ref{fig:3d_n2q_fid_arb}}. Here, we generate random circuits with 40 qubits and varying average two-qubit gates per qubit and qubit degrees. Two-qubit gate per qubit indicates how many gates involve a certain qubit. The degree is the number of unique qubits with which a given qubit interacts. The two-qubit gate count informs the circuit depth, whereas the degree indicates the gate locality. In the figure, each surface's X and Y axes represent the two-qubit gate per qubit and qubit degree, respectively. The Z axis is the number of two-qubit gates and the relative fidelity improvement for the first and second rows. In the second row, we also project the surface into plane A (the Y-Z plane colored green) and plane B (the X-Z plane colored blue). Each line in plane A represents the fidelity improvement under a fixed two-qubit gate per qubit and varying degrees. In contrast, each line in plane B represents the fidelity improvement under a fixed degree and varying two-qubit gate per qubit. This setting is similar in Fig.~\ref{fig:3d_n2q_fid_qaoa} and Fig.~\ref{fig:3d_n2q_fid_qsim}.
The results show the following insights clearly:

(1) \mbox{\name} excels in \textbf{high-degree circuits}. In low-degree, well-localized circuits, \name's performance falls marginally behind FAA due to higher decoherence incurred by movement.
(2) \textbf{Deeper circuits benefit more from \name}, evidenced by widening fidelity gaps as the two-qubit gate count per qubit increases.

\begin{table}[t]
\centering
\renewcommand*{\arraystretch}{1}
\setlength{\tabcolsep}{2pt}

\caption{Number of multi-qubit pulses (lower the better)}

\begin{tabular}{l|ccccc}
\toprule
Benchmark & HHL-7 & Mermin-Bell-10 & QV-32 & BV-50 & BV-70 \\
\midrule
Geyser~\cite{geyser} & 486 & 564 & 11803 & 432 & 655 \\
\textbf{\name} & \textbf{348} & \textbf{306} & \textbf{4995} & \textbf{66} & \textbf{108} \\
\bottomrule
\end{tabular}

\label{tab:comp_geyser}
\vspace{-15pt}

\end{table}

\textbf{(b) QAOA and Quantum Simulation Circuits.} In Fig.~\ref{fig:3d_n2q_fid_qaoa} and \ref{fig:3d_n2q_fid_qsim}, we present results for QAOA and QSim circuits with varying properties. For QAOA, we generate regular graphs with different degrees. For QSim, we change the probability for each Hamiltonian term being a non-$I$ operator. The graph degree in QAOA and probability of non-I terms resemble the qubit degree discussed in the previous paragraph, as they all indicate the locality of the problem.
The previous insights are still valid here. The more non-local the program is, the higher the advantage \name is. In addition, \name is better for circuits with more qubits and depth.

\noindent\textbf{Sensitivity Analysis on Hardware Parameters.} In Fig.~\ref{fig:ablation_2d_all}, we conduct a sensitivity analysis on six key hardware parameters. The top row displays the circuit fidelities for three benchmarks—BV-70, QSim-rand-20, and QAOA-regu5-40 under different parameter settings. Comparative data for \name, FAA-Rectangular, and FAA-Triangular are also provided. The second row elucidates the error breakdown for BV-70 by calculating $-\log(\text{fidelity})$. Various error sources are color-coded, including one-qubit gate error, two-qubit gate error, movement heating error, cooling overhead error, atom loss during movement, and movement decoherence error.

\textbf{(a) Time per Move.} We vary the time per movement from 100 to 1000$\upmu$s. According to Eq.~\mbox{\ref{eq:fheat}}, shorter times increase the vibrational quantum number ($n_\text{vib}$), causing significant atom loss. In the extreme case, where T\_per\_move smaller than 150$\upmu$, the heating of one single step can exceed $N_{max}$, making the cooling procedure no use. When T\_per\_move is smaller than 200$\upmu$, the cooling can help reduce movement heating error but inevitably incurs large cooling overhead. As T\_per\_move increases, the decoherence error dominates. The optimal setting is around 300$\upmu$s, where the decoherence and heating are comparable, \textit{within realistic 200-400$\upmu$s range}~\cite{bluvstein2022quantum}.

\textbf{(b) Average Move Speed.} We also present the previous results in a different way by showing the atom movement speed, which is inversely related to T\_per\_move. Faster speeds induce atom loss, while slower speeds elevate decoherence error. QAOA and QSim benchmarks exhibit faster optimal speeds than arbitrary circuits, attributable to their smaller size.

\textbf{(c) Atom Distance.} We vary the SLM atom separation from 1$\upmu$m to 60$\upmu$m. Larger distances proportionally elevate $n_\text{vib}$ with $D^2$ rate, increasing heating error. Cooling operations stabilize this error but introduce overhead, dominating at 60$\upmu$m. \name outperforms baselines for distances under 40$\upmu$m. In experiments, we adopt the 15$\upmu$m setting from \cite{bluvstein2022quantum}, where the heating error is minimal.

\textbf{(d) $n_\text{vib}$ for Cooling.} We employ cooling to mitigate the heating and atom loss caused by excessive $n_\text{vib}$. Frequent cooling introduces two-qubit gate errors. Using a 60$\upmu$s atom distance to examine trade-offs, we find a low threshold causes significant cooling errors, while a high threshold results in severe atom loss. \textit{An optimal threshold range of 12 to 25 minimizes overall impact}; we use 15 in our experiments.

\textbf{(e) Coherence Time.} Coherence time impacts more on \mbox{\raa} than to FAA since \mbox{\raa} execution time is much longer due to movements. With movement time (300$\upmu$s) significantly exceeding one-qubit and two-qubit gate times (625ns, 380ns), \mbox{\raa} benefits more from extended coherence time. Comparative analysis shows \mbox{\raa} outperforms when coherence time exceeds 1s, a threshold generally met in practice as per~\mbox{\cite{bluvstein2022quantum}} and~\mbox{\cite{madjarov2020high}}.

\textbf{(f) two-qubit Gate Fidelity.} For fidelities below 0.9999, two-qubit gate errors dominate, as indicated in (f). Remarkably, FAA and Geyser outperform \mbox{\raa} when two-qubit fidelity exceeds 0.9999, owing to minimal SWAP overheads. Nonetheless, under current achievable two-qubit fidelity and coherence time~\cite{bluvstein2022quantum,evered2023high} \name remains more reliable.

\begin{figure}[t]
    \centering
\subfloat[Same number of qubits, different row column ratio (shape).]{%
  \includegraphics[clip,width=\columnwidth]{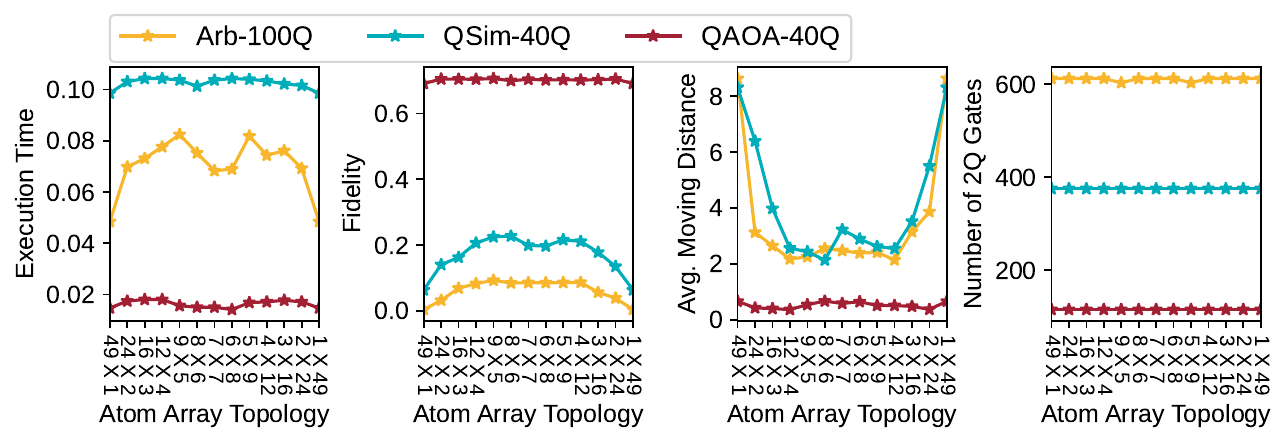}%
    }    
    \vspace{-10pt}
\subfloat[Different number of qubits, number of rows = number of columns]{%
  \includegraphics[clip,width=\columnwidth]{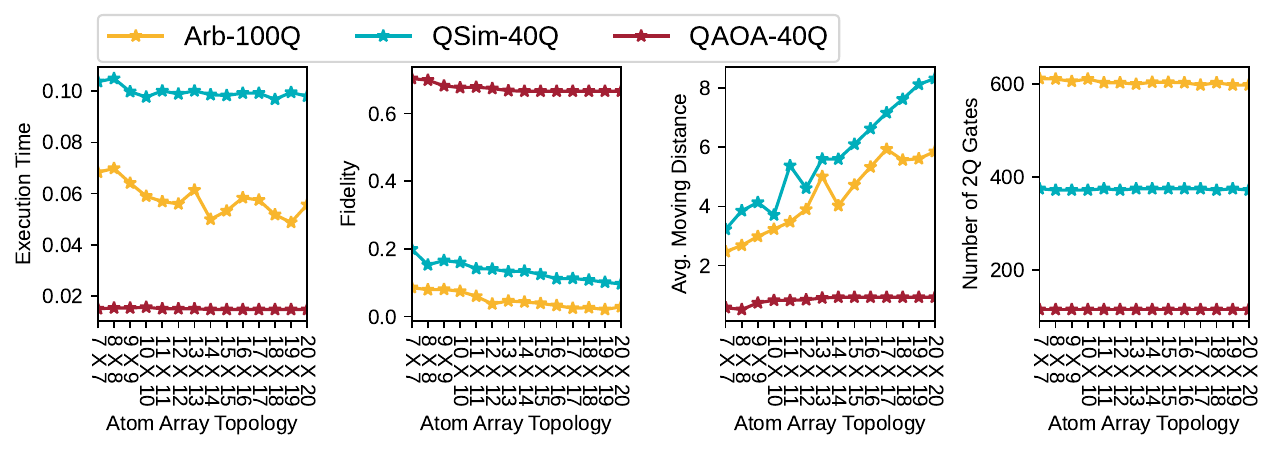}%
    }  
    \vspace{-10pt}
\subfloat[Different number of AOD arrays]{%
  \includegraphics[clip,width=\columnwidth]{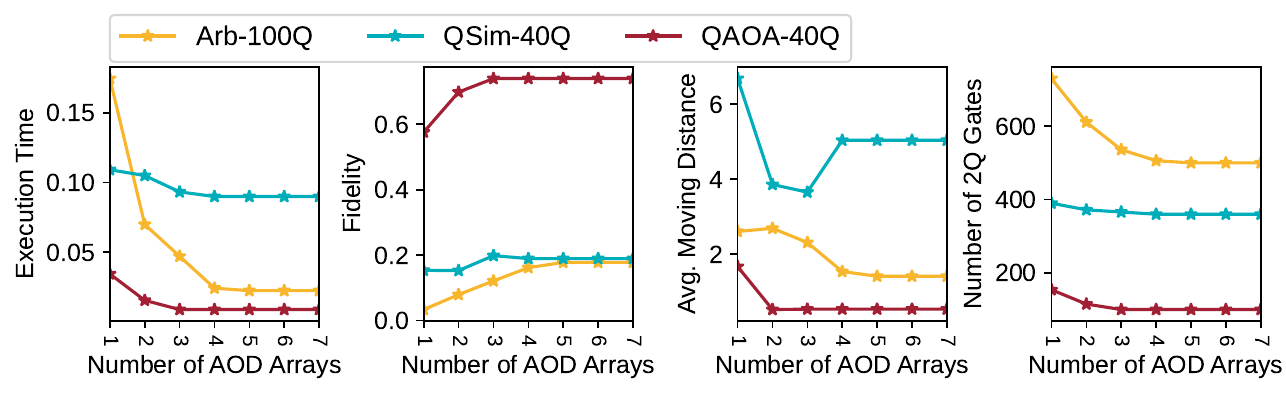}%
    }

    \vspace{-5pt}
    \caption{Sensitivity analysis of array topology.}

    \label{fig:dev_ablation_2d_n_rows_same_size}
    \vspace{-17pt}
\end{figure}

\noindent\textbf{Impact of the Array Topology.}
We also examine the execution time, fidelity, average moving distance (mm), and two-qubit gate count under different array configurations. Benchmarks are 100Q arbitrary circuit with ten gates per qubit; 40Q simulation with $p=0.5$ probability being non-$I$ and ten strings; 40Q QAOA with 5-regular graph. 

\textbf{(a) Different Row-Col Ratio under Same Qubit Number}
Fig.~\ref{fig:dev_ablation_2d_n_rows_same_size}(a) examines the impact of array shape while maintaining the same overall atom count. For all three tasks, square arrays maximize fidelity due to shorter movement distances, as the distance plot shows. However, the execution time increases slightly for square-like topology because they impose stronger constraints for the movement and reduce the parallelism.

\textbf{(b) Different Qubit Number, Row=Col.} Fig.~\ref{fig:dev_ablation_2d_n_rows_same_size}(b) analyzes the effects of varying array sizes from $7\times7$ to $20\times20$. The optimal fidelity is achieved for all three tasks with a $7\times7$ array. As the array size expands, \mbox{\name} preferentially maps qubits to diagonal atoms, increasing parallelism and reducing execution time. However, this diagonal mapping lengthens moving distances without significantly altering the two-qubit gate count, leading to decreased fidelity due to \textit{heating}. 

\textbf{(c) Different Number of AOD Arrays.}
Fig.~\ref{fig:dev_ablation_2d_n_rows_same_size}(c) examines the impact of the number of AODs, ranging from 1 to 7. Additional AODs enhance the coupling map, and the relaxation of order-preserving constraints allows for more efficient movements—these two reasons combined reduce the two-qubit gate count and execution time. The smaller gate count and shorter execution time help increase the fidelity.

\begin{figure}[t]
    \centering
    \includegraphics[width=\columnwidth]{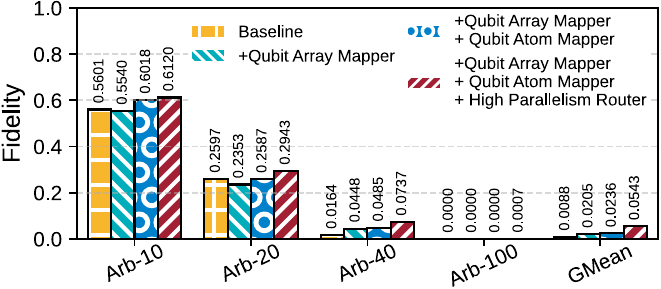}
    \vspace{-15pt}
    \caption{Breakdown of improvement of compiler techniques.}
    \label{fig:technique_breakdown}
    \vspace{-15pt}
\end{figure}
\begin{figure}[t]
    \centering
    \includegraphics[width=\columnwidth]{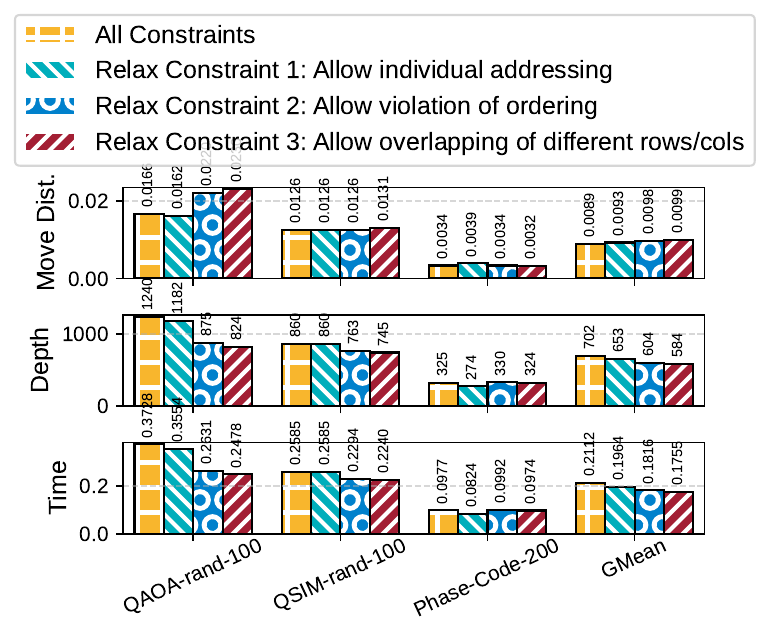}
    \vspace{-20pt}
    \caption{Performance of \mbox{\name} after relaxing one of the three constraints.}
    \label{fig:ablation_contraints_release}
    \vspace{-15pt}
\end{figure}

\noindent\textbf{Breakdown of Technique-Induced Improvements.} Fig.~\ref{fig:technique_breakdown} delineates the contributions of individual techniques to overall performance. We evaluated random circuits with 26 gates per qubit. For the baseline, we adopt the qiskit's dense mapping instead of the greedy mapping that maximizes the inter-array cut; we also use a random qubit-atom mapper rather than the load balance mapper; finally, we use a serial router that performs one gate at a time instead of parallelizing the gate. For the other experiments, we add these techniques cumulatively. On average, our qubit-array mapper gives rise to 3.53$\times$ better fidelity than the dense mapper and subsequent diagonal-first qubit-atom mapper gives rise to 1.19$\times$ higher fidelity. Finally, the high parallelism router brings 2.59$\times$ fidelity improvement. Combining all three methods, we have \textbf{10.9$\times$} higher fidelity.

\noindent\textbf{Relaxation of Hardware Constraints.}
Our compiler also enables toggling the constraints associated with moving AODs. Fig.~\mbox{\ref{fig:ablation_contraints_release}} presents the outcomes of loosening the three constraints detailed in Fig.~\mbox{\ref{fig:router_checks}}, \mbox{\ref{fig:router_checks1}}, and \mbox{\ref{fig:router_checks2}}. The count of two-qubit gates remains unchanged, as these constraints only influence the scheduling of the two-qubit gates. The depth is reduced, attributed to the increased flexibility in gate scheduling, which concurrently lowers the execution time. Conversely, the average movement distance sees an uptick due to the enhanced movement flexibility. The results indicate that among the constraints, easing the third yields the most significant improvement in performance.

\begin{figure}[t]
    \centering
    \includegraphics[width=\columnwidth]{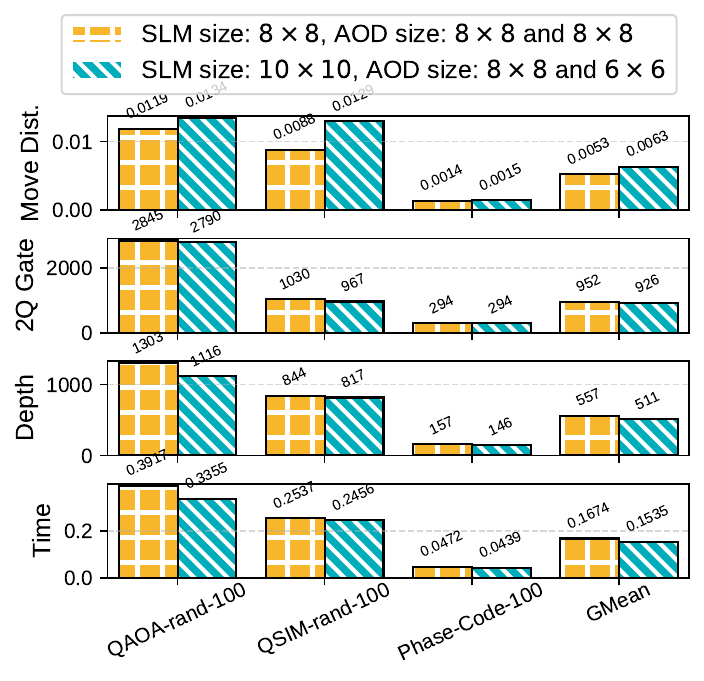}
    \vspace{-15pt}
    \caption{Performance with different sizes across AOD layers.}
    \label{fig:ablation_different_aod_size}
    \vspace{-15pt}
\end{figure}

\noindent\textbf{Variability in AOD sizes.}
Our compiler accommodates AODs featuring distinct sizes across layers. In Fig.\mbox{~\ref{fig:ablation_different_aod_size}}, we showcase outcomes for two configurations: one with uniform AOD and SLM dimensions, and another with varying dimensions. The disparity in SLM and AOD dimensions facilitates a reduction in the number of two-qubit gates as well as a decrease in both depth and execution time, as the varied sizes provide the compiler with enhanced flexibility in qubit mapping. Nonetheless, an increase in the moving distance is an inevitable consequence of the enlarged SLM dimensions.

\begin{figure}[t]
    \centering
    \includegraphics[width=\columnwidth]{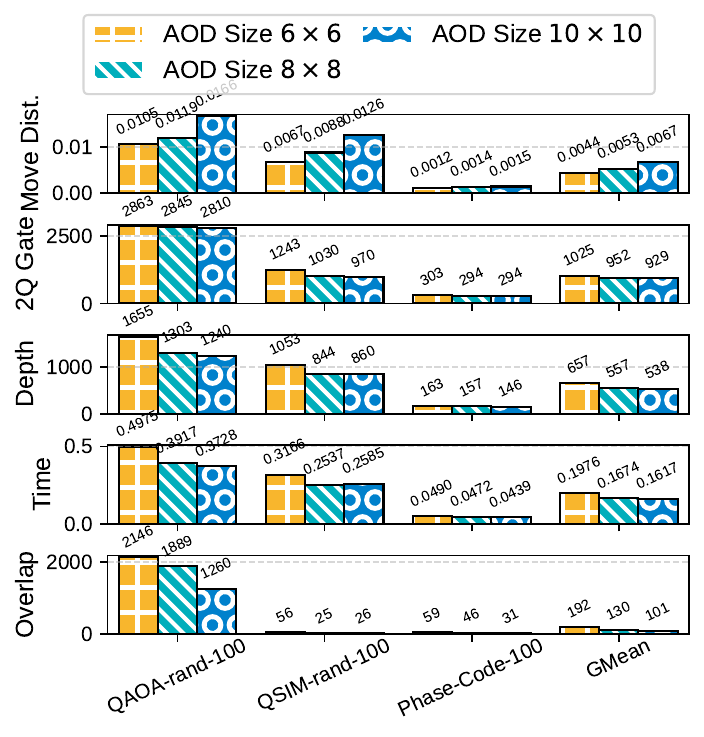}
    \vspace{-20pt}
    \caption{Overlap and other quantities when the number of logical qubits is close to that of physical qubits.}
    \label{fig:ablation_logical_close_physical}
    \vspace{-15pt}
\end{figure}

\noindent\textbf{Overlapping under Extreme Conditions.} We analyze the performance when the number of logical qubits required by the circuit closely approaches the number of qubits available in the hardware. In Fig.~\mbox{\ref{fig:ablation_logical_close_physical}}, we compile circuits with 100 qubits on hardware where each AOD or SLM ranges in size from $6\times 6$ (108 physical qubits in total) to $10\times 10$ (300 physical qubits in total). Aside from an expected decrease in two-qubit gate depth and execution time, as well as an increase in moving distance, we also record the number of overlaps during compilation. This is defined as the case when the compiler fails to insert more two-qubit gates into the same layer due to a violation of Constraint 3. From the figure, it is evident that increasing the AOD size can aid in reducing overlaps, thereby optimizing gate scheduling. Furthermore, the overlapping behavior is highly application-dependent.

\begin{figure}[t]
    \centering
    \includegraphics[width=1\columnwidth]{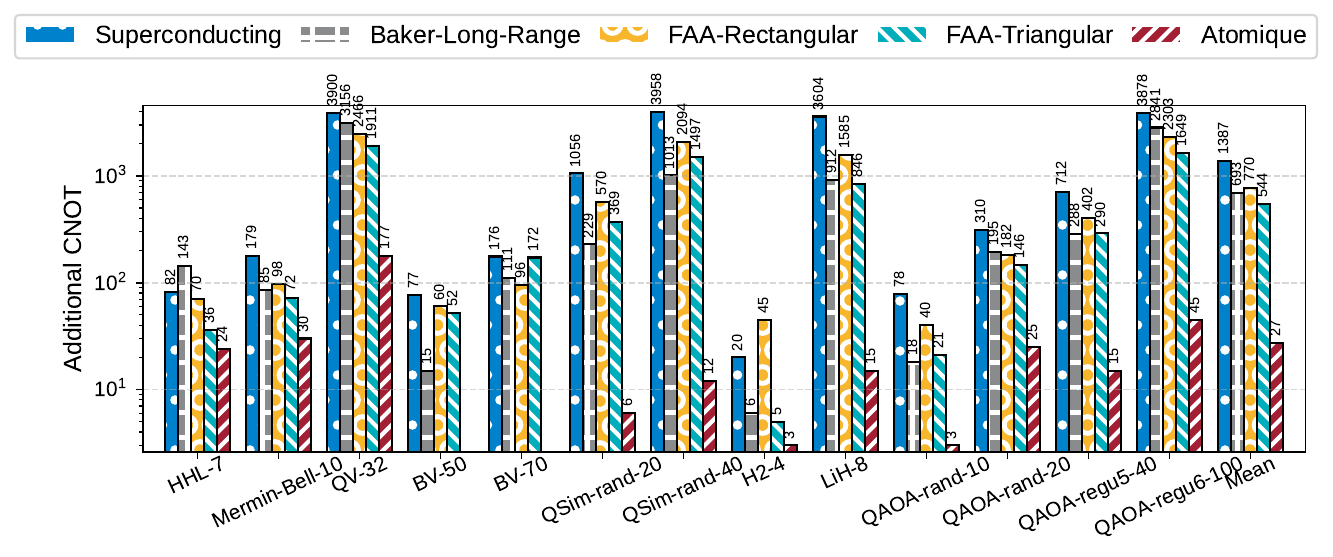}
    \vspace{-16pt}
    \caption{Comparisons of additional \cnot\mbox{\xspace} caused by SWAP-insertion among different architectures.}
    \label{fig:swap_count}
    \vspace{-20pt}
\end{figure}

\noindent\textbf{Additional \cnot\mbox{\xspace} Due to SWAP Insertion.}
We record the number of \cnot\mbox{\xspace} gates added due to SWAP insertion during compilation, which is the difference in the count of \cnot s\mbox{\xspace} before and after this step. From Fig.\mbox{~\ref{fig:swap_count}}, it is evident that the additional \cnot s\mbox{\xspace} when using \mbox{\name} are consistently lower than those of all baselines. This performance is particularly notable in benchmarks such as BV and QSim.

\section{Related Works}

\textbf{Compilers for Emerging Quantum Architectures}
In addition to widely used superconducting quantum computers, researchers have recently shown increasing interest in new quantum computing hardwares such as neutral-atom machines and trapped-ion machines. In this work, we have discussed the previous work design for neutral-atom machines~\cite{baker2021exploiting, geyser, tan2022}. In addition, a number of compilers designed for trapped ion systems have been proposed in recent years, including TILT~\cite{wu2020tilt} for a linear chain of trapped ions, Ref.~\cite{murali2020} for quantum charge coupled device-based trapped ion architectures, and Ref.~\cite{Kreppel:2022oyr} for shuttling-based trapped ion architectures. Many compilation techniques for various settings have also been proposed~\cite{ravi2022vaqem, 10.1145/3503222.3507703, das2021jigsaw, 10.1145/3445814.3446743,murali2019noise, tannu2019not, li2019tackling, molavi2022qubit, liuqucloud, zhang2021time,merrill2014progress, low2014optimal, brown2004arbitrarily, xie2022suppressing, hahn1950spin, viola1999dynamical, biercuk2009optimized, lidar2014review, das2021adapt,wallman2016noise,zhang2021hidden,murali2020software, wu2020tilt,versluis2017scalable, helmer2009cavity, ding2020systematic,li2022paulihedral, lao20222qan, cheng2022topgen}. In this work, we focus on the emerging \raa devices implemented with dynamically reconfigurable atom arrays, which require new techniques to fully exploit their power. Therefore, we design a compilation framework that delivers low-depth compiled circuits with high scalability.

\textbf{Qubit Mapping and Instruction Scheduling}
Since noise is the bottleneck of NISQ machines, many noise-adaptive quantum compilation techniques have been proposed~\cite{ravi2022vaqem, 10.1145/3503222.3507703, das2021jigsaw, 10.1145/3445814.3446743, hua+:arxiv_crosstalk}. Examples include various gate errors which can be suppressed by qubit mapping~\cite{murali2019noise, tannu2019not, li2019tackling, molavi2022qubit, liuqucloud, zhang2021time}, composite pulses~\cite{merrill2014progress, low2014optimal, brown2004arbitrarily, xie2022suppressing}, dynamical decoupling ~\cite{hahn1950spin, viola1999dynamical, biercuk2009optimized, lidar2014review, das2021adapt}, randomized compiling~\cite{wallman2016noise}, hidden inverses~\cite{zhang2021hidden}, instruction scheduling~\cite{murali2020software, wu2020tilt}, frequency tuning~\cite{versluis2017scalable, helmer2009cavity, ding2020systematic}, parallel execution on multiple machines~\cite{stein2022eqc}, algorithm-aware compilation~\cite{jin+:asplos23_qaoa, jin:arxiv2024_tetris, jin:arxiv2024_qft, li2022paulihedral, lao20222qan, cheng2022topgen}, qubit-specific basis gate~\cite{lin2022let, li2021software}, and vairous pulse level optimizations~\cite{chen+:hpca23_pulse}. 
Qubit mapping and routing, also named quantum layout synthesis or qubit allocation/placement, has been a popular research topic in the quantum computing community \cite{li2019tackling, dac19-wille-burgholzer-zulehner-mapping-minimal-swaph, isca19-murali-linke-martonisi-abhari-nguyen-alderete-triq-architecture-studies, zhang2021time, iccad21-tan-cong-qubit-mapping-absorption, iccad20-tan-cong-optimal-layout-synthesis, tcad08-maslov-falconer-mosca-placement, cgo18-siraichi-santos-collange-pereira-qubit-allocation, date18-zulehner-paler-wille-efficient-mapping-ibmqx, zhou_monte_2020, molavi2022qubit, fan-reinforcement-learning}. Instruction scheduling has also been widely explored~\cite{murali2020software, wu2020tilt, smith2021, stein2022eqc,lin2022let, li2021software}. The most relevant previous work is Brandhofer \etal~\cite{iccad21-brandhoher-buchler-polian-optimal-mapping-atoms} which considers a hypothetical atom array architecture with ``1D displacement'' reconfigurability, a much more restrictive architecture than the \raa we discussed in this paper.

\section{Conclusion and Outlook}
We architect the emerging \raafull (\raa) with moveable atoms at runtime.
In this highly flexible \raa, we propose a qubit mapper and a router to reduce the SWAP overhead and maintain high parallelism of two-qubit gate executions. Compared to other architectures or compilers, our \name significantly reduces the depth and two-qubit gates.
We hope \name opens up an avenue for future \raa.

\section{Acknowledgement}
This work is partially supported by NSF grant 2313083, CCF-1901381, CCF-2115104, CCF-2119352, and CCF-2107241, MIT-IBM Watson AI Lab, and Qualcomm Innovation Fellowship. The authors would like to thank Dolev Bluvstein, Mikhail D. Lukin, and Hengyun Zhou for valuable discussions on neutral atom arrays.

\bibliographystyle{IEEEtranS}
\bibliography{main} 

\newpage

\appendix
\section{Artifact Appendix}

\subsection{Abstract}

Our paper introduces a new compiler, \name, which is tailored for qubit mapping, atom movement, and gate scheduling of the \raas.

The artifact contains the \name compiler, which is this paper's main contribution, baseline compilers used for comparison, benchmarks, and necessary scripts to run experiments and draw figures. 

\subsection{Artifact check-list (meta-information)}

{\small
\begin{itemize}
  \item {\bf Algorithm: }
  The compiler is mainly built from three components, which are our original algorithms:
  \begin{itemize}
      \item Qubit Array Mapper.
      \item Qubit Atom Mapper.
      \item High Parallelism Router.
  \end{itemize}
  \item {\bf Program: }
  We use a variety of benchmarks to test our compiler, including:
  \begin{itemize}
      \item Algorithm circuit from QASMBench.
      \item Algorithm circuit from SupermarQ.
      \item QAOA circuit generated by us.
      \item QSim circuit generated by us.
  \end{itemize}
  All are public. We have already included them in the artifact for ease of use.
  \item {\bf Compilation: }
    Python 3.9 with Qiskit 0.38.0. All are public-available.
  \item {\bf Run-time environment: }
    Python with Qiskit and Pytorch. We tested it only on Linux; however, it should be able to run on any platform that supports Python, Qiskit, and Pytorch.
  \item {\bf Hardware: }
  There is no specific hardware requirement. To complete the experiments in a reasonable amount of time, a CPU with 8 cores and 16 GB of memory is recommended.
  \item {\bf Run-time state: }
  \begin{itemize}
    \item Not OS-speciﬁc.
    \item No need for root access.
    \item Not run-time sensitive.
  \end{itemize}

  \item {\bf Execution: }
    \begin{itemize}
    \item No specific condition is required.
    \item The data for each figure can be generated in less than 10 minutes, except Tab.3. The baseline Geyser takes 10 hours to a day to complete.
  \end{itemize}
  \item {\bf Metrics: }
  \begin{itemize}
      \item two-qubits gate count.
      \item one-qubit gate count.
      \item Fidelity.
      \item Circuit depth.
      \item Compilation time.
  \end{itemize}
  In general, the generated figures should match those presented in the paper.
  \item {\bf Output: }
Figures 12-26 of the article and the necessary data for Tables 2 and 3.
  \item {\bf Experiments: }
  Please see Readme.md in the artifact for more detailed instructions on running the experiments.
  \item {\bf How much disk space required (approximately)?: }
  $<$10 GB
  \item {\bf How much time is needed to prepare workflow (approximately)?: }
  $<$10 minutes
  \item {\bf How much time is needed to complete experiments (approximately)?: }
  $<$2 hours if excluding the Geyser baseline.
  \item {\bf Publicly available?: }
  Our compiler is available at \url{https://doi.org/10.5281/zenodo.10895509}.
  \item {\bf Code licenses (if publicly available)?: }
MIT license.
  \item {\bf Workflow framework used?: }
  Qiskit and PyTorch.
  \item {\bf Archived (provide DOI)?: }
  \url{https://doi.org/10.5281/zenodo.10895509}
\end{itemize}
}

\subsection{Description}

\subsubsection{How to access}

The artifact is available at the following link:
\url{https://doi.org/10.5281/zenodo.10895509}

\subsubsection{Hardware dependencies}
To complete the experiments in a reasonable amount of time, a CPU with 8 cores and 16 GB of memory is recommended.
\subsubsection{Software dependencies}
The artifact is implemented in Python and requires several packages, such as PyTorch and Qiskit. The complete list of required packages is in requirements.txt in the artifact folder.
\subsubsection{Data sets}
All the benchmarks are provided in the artifact.

\subsection{Installation}

\begin{enumerate}
    \item \textbf{Install Required Packages}: Assuming you are using Anaconda to manage the Python environment, install all the necessary Python packages listed in the \texttt{requirements.txt} file using the following command:
    
    \begin{lstlisting}[language=bash]
    conda create -n atomique
    conda activate atomique
    conda install python==3.9
    pip install -r requirements.txt
    \end{lstlisting}
    
    \item \textbf{Set Up Environment Variable}: Export the Python path to the base directory to ensure the scripts run correctly.
    
    \begin{lstlisting}[language=bash]
    export PYTHONPATH=.
    \end{lstlisting}
\end{enumerate}

\subsection{Experiment workflow}
To run all the experiments, run
    \begin{lstlisting}[language=bash]
    ./all_script
    \end{lstlisting}
\subsection{Evaluation and expected results}

To evaluate the functionality of \name, the quantum circuit should be successfully compiled into basic operations supported by \raa, and some statistics of the compiled circuit should also be output. 

For the evaluation of \name, all the figures presented in the paper should be reproduced using the artifact. The figures are stored under the plot folder after running the experiments.

\subsection{Experiment customization}
The user can write their own benchmarks and configs to customize the experiments by following the provided example benchmarks and configs.

\subsection{Methodology}

Submission, reviewing and badging methodology:

\begin{itemize}
  \item \url{https://www.acm.org/publications/policies/artifact-review-and-badging-current}
  \item \url{http://cTuning.org/ae/submission-20201122.html}
  \item \url{http://cTuning.org/ae/reviewing-20201122.html}
\end{itemize}
\end{document}